\documentclass[11pt,a4paper]{article}

\usepackage{subfigure}
\usepackage{amsthm,amsmath}
\usepackage[round]{natbib}
\usepackage{graphicx}
\usepackage{bm,amssymb}
\usepackage{algpseudocode,algorithm}
\RequirePackage[colorlinks,citecolor=blue,urlcolor=blue]{hyperref}
\usepackage[font=small,labelfont=bf]{caption}

\algnewcommand\algorithmicto{\textbf{to}}

\newcommand{\pinv}[1]{#1^{\dagger}}
\newcommand{\ownint}[4]{{\int_{#1}^{#2} \! #3 \, \mathrm{d}#4}}
\newcommand{\tp}[1]{{#1^{\mathrm{T}}}}
\newcommand{\E}{\operatorname{E}}
\DeclareMathOperator*{\argmax}{arg\,max\,}

\begin{document}

\title{Empirical Bayes unfolding of elementary particle spectra at the Large Hadron~Collider}
\author{Mikael Kuusela\thanks{Supported in part by a grant from the Helsinki Institute of Physics.} \\ \texttt{mikael.kuusela@epfl.ch} \and Victor M. Panaretos \\ \texttt{victor.panaretos@epfl.ch}}
\date{}
\maketitle
\vspace{-1cm}
\begin{center}
Section de Math\'{e}matiques \\
\'{E}cole Polytechnique F\'{e}d\'{e}rale de Lausanne \\
EPFL Station 8, 1015 Lausanne \\
Switzerland
\end{center}

\begin{abstract}
We consider the so-called unfolding problem in experimental high energy physics, where the goal is to estimate the true spectrum of elementary particles given observations distorted by measurement error due to the limited resolution of a particle detector. This an important statistical inverse problem arising in the analysis of data at the Large Hadron Collider at CERN. Mathematically, the problem is formalized as one of estimating the intensity function of an indirectly observed Poisson point process. Particle physicists are particularly keen on unfolding methods that feature a principled way of choosing the regularization strength and allow for the quantification of the uncertainty inherent in the solution. Though there are many approaches that have been considered by experimental physicists, it can be argued that few -- if any -- of these deal with these two key issues in a satisfactory manner. In this paper, we propose to attack the unfolding problem within the framework of empirical Bayes estimation: we consider Bayes estimators of the coefficients of a basis expansion of the unknown intensity, using a regularizing prior; and employ a Monte Carlo expectation-maximization algorithm to find the marginal maximum likelihood estimate of the hyperparameter controlling the strength of the regularization. Due to the data-driven choice of the hyperparameter, credible intervals derived using the empirical Bayes posterior lose their subjective Bayesian interpretation. Since the properties and meaning of such intervals are poorly understood, we explore instead the use of bootstrap resampling for constructing purely frequentist confidence bands for the true intensity. The performance of the proposed methodology is demonstrated using both simulations and real data from the Large Hadron Collider.

\end{abstract}

\section{Introduction} \label{sec:intro}

This paper studies a generalized linear inverse problem \citep{Bochkina2013}, called the {\em unfolding problem} \citep{Prosper2011, Cowan1998, Blobel2013}, arising in the analysis of the data produced at the Large Hadron Collider (LHC) at CERN, the European Organization for Nuclear Research. The LHC is the world's largest and most powerful particle accelerator. It collides two beams of protons in order to study the properties and interactions of elementary particles produced in such collisions. The trajectories and energies of these particles are recorded using four gigantic underground particle detectors and the vast amounts of data produced by these experiments are analyzed in order to draw conclusions about fundamental laws of physics. Due to their complex structure and huge quantity, the analysis of these data poses significant statistical and computational challenges.

Physicists use the term ``unfolding" to refer to correcting the distributions measured at the LHC for the limited resolution of the particle detectors. Let $X$ be some physical quantity of interest measured in the detector. This could, e.g., be the energy, mass or production angle of a particle. Due to the noise induced by the detector, we are only able to observe a stochastically {\em smeared} or {\em folded} version $Y$ of this quantity. As a result, the observed distribution of $Y$ is a ``blurred'' version of the true, physical distribution of $X$ and the task is to use the observed values of $Y$ to estimate the distribution of~$X$.

The main challenge in unfolding is the ill-posedness of the problem in the sense that a simple inversion of the forward mapping from the true space into the smeared space is unstable with respect to small perturbations of the data \citep{Engl2000,Kaipio2005, Panaretos2011}. As such, the trivial maximum likelihood solution of the problem often exhibits spurious high-frequency oscillations. These oscillations can be tamed by regularizing the problem which is done by taking advantage of additional a priori knowledge about plausible solutions.

An additional complication is the non-Gaussianity of the data which follows from the fact that both the true and the smeared observations are realizations of two interrelated Poisson point processes denoted by $M$ and $N$, respectively. As such, unfolding is an example of a {\em Poisson inverse problem} \citep{Antoniadis2006, Reiss1993} where the intensity function $f$ of the true process $M$ is related to the intensity function $g$ of the smeared process $N$ via a Fredholm integral operator $K$; that is, $g = Kf$, where $K$ represent the response of the detector. The task at hand is then to estimate and make inferences about the true intensity $f$ given a single observation of the smeared process $N$. Due to the Poisson nature of the data, many standard techniques based on a Gaussian likelihood, such as Tikhonov regularization, are only approximately valid for unfolding. Furthermore, estimators properly taking into account the Poisson distribution of the observations are rarely available in a closed form making the problem computationally challenging.

At present, the unfolding methodology used in LHC data analysis is far from being well-established \citep{Lyons2011}. The two main approaches are the expectation-maximization (EM) algorithm with an early stopping \citep{DAgostini1995,Vardi1985,Lucy1974,Richardson1972}, and a certain variant of Tikhonov regularization \citep{Hoecker1996}. In high energy physics (HEP) terminology, the former is called the \emph{D'Agostini iteration} and the latter, somewhat misleadingly, \emph{SVD unfolding} (with SVD standing for singular value decomposition). In addition, a HEP-specific heuristic, called \emph{bin-by-bin unfolding}, which provably accounts for smearing effects incorrectly through a multiplicative efficiency correction, is widely used. Recently, \citet{Choudalakis2012} proposed a Bayesian solution to the problem, but this seems to have seldom been used in practice, thus far.

The main problem with the D'Agostini iteration is that it is difficult to give a physical interpretation to the regularization imposed by early stopping of the iteration. SVD unfolding, on the other hand, ignores the Poisson nature of the observations and does not enforce the positivity of the solution. Furthermore, both of these methods suffer from not dealing with two significant issues satisfactorily: (1) the choice of the regularization strength and (2) quantification of the uncertainty in the solution. The delicate problem of choosing the regularization strength is handled in most LHC analyses using non-standard heuristics or, in the worst case scenario, by simply fixing a certain value ``by hand". When quantifying the uncertainty of the unfolded spectrum, the analyses rarely attempt to take into account the uncertainty related to the choice of this regularization strength. Each year, the experimental collaborations working with LHC data publish dozens of papers using such unsatisfactory unfolding techniques. Recent examples include studies of the characteristics of jets \citep{CMS2012Jets}, the transverse momentum distribution of $W$ bosons \citep{ATLAS2012W} and charge asymmetry in top-quark pair production \citep{CMS2012Top}, to name a few.

In this paper, we propose a novel unfolding technique aimed at addressing the above-mentioned issues within a principled framework. The main features of our method, which casts the problem as Bayesian estimation of series expansion coefficients of the intensity, subject to a regularising prior, are:
\begin{itemize}
\item Empirical Bayes selection of the regularization parameter using a Monte Carlo expectation-maximization algorithm \citep{Geman1985, Geman1987, Saquib1998, Casella2001};
\item Frequentist uncertainty quantification, including the uncertainty of the regularization parameter, using the parametric bootstrap.
\end{itemize}
To the best of our knowledge, neither of these techniques has been previously used to solve the HEP unfolding problem. Our method also properly takes into account the Poisson distribution of the observations, enforces the positivity constraint of the unfolded spectrum and imposes a curvature penalty on the solution with a straightforward physical interpretation.

The unfolding problem is closely related to image reconstruction in emission tomography \citep{Shepp1982, Vardi1985, Green1990} and to image deblurring in optics \citep{Richardson1972} and astronomy \citep{Lucy1974} --- once discretized, all of these are described by a similar Poisson regression problem. There are however at least three important differences between these problems and unfolding. First, in tomography and image processing, the unknown is a two- or three-dimensional image, while in HEP unfolding one is typically interested in a one-dimensional intensity spectrum. This makes the scale of the problem at least an order of magnitude smaller enabling the use of computationally intensive statistical methods, such as the ones described in this paper. Second, uncertainty quantification of the solution is crucial in high energy physics which is rarely the case on other domains using similar models; and third, images are in principle naturally discretized using pixels, while for HEP spectra other basis expansions can be more appropriate.

Classical, well-understood techniques for choosing the regularization \linebreak strength in inverse problems include the Morozov discrepancy principle \citep{Morozov1966} and cross-validation \citep{Stone1974}. \citet{Bradsley2009} study these techniques in the context of Poisson inverse problems, while \citet{Veklerov1987} provide an alternative approach based on statistical hypothesis testing. On the contrary, empirical Bayes selection of the regularization parameter, one of the key elements of our unfolding procedure, has received relatively less attention in the literature. Among the few recent contributions, \citet{Johnstone2005} demonstrated the good performance of the marginal maximum likelihood estimator (MMLE) in choosing the threshold levels in wavelet smoothing for direct observations under Gaussian noise. The approach we follow bears similarities to that of \citet{Saquib1998} where the MMLE is used to select the regularization parameter in tomographic image reconstruction with Poisson data. In spite of their demonstrated good performance on many real-world datasets, empirical Bayes techniques have not become widely-used in tomography due to their high computational cost \citep{Leahy2000,Green2012}. In our case, however, the smaller scale of the problem makes the computations tractable on a modern desktop computer.

The second key element of our methodology is frequentist uncertainty quantification based on the parametric bootstrap. The use of the credible intervals of the empirical Bayes posterior would provide the most straightforward way of giving confidence statements for this problem, but due to the data-driven choice of the hyperparameter, these intervals do not enjoy the same subjective interpretation as standard Bayesian intervals. Moreover, in HEP, frequentist confidence statements are generally preferred over Bayesian uncertainty quantification \citep{Lyons2013}. For these reasons, we explore the use of simple, albeit computationally expensive, bootstrap resampling for constructing frequentist confidence bands for the unknown intensity. This also enables us to take into account the uncertainty regarding the choice of the regularization parameter which is usually ignored in related frequentist procedures \citep{Berk2013, Efron2013}. A sensible alternative to our methodology would be to use hierarchical Bayes by placing a hyperprior on the unknown regularization parameter \citep{Kaipio2005}. Such an approach would enable an automatic choice of the regularization strength along with standard Bayesian uncertainty quantification, but is dependent on the choice of the hyperprior. In effect, our proposed methodology carries over the benefits of hierarchical Bayes to the frequentist setting without the need to worry about the choice of the hyperprior.

The paper is structured as follows. Section \ref{sec:physics} provides the necessary background on the experimental data produced at the LHC and the role of unfolding in the analysis of these data. We then formulate in Section \ref{sec:formulation} a forward model for the unfolding problem using Poisson point processes. The proposed methodology of empirical Bayes unfolding is explained in detail in Section \ref{sec:EBU} which forms the backbone of this paper. This is followed by simulation studies in Section \ref{sec:simulations} and a real-world data analysis scenario in Section \ref{sec:Zboson} consisting of the unfolding of the $Z$ boson invariant mass spectrum measured at the CMS experiment at the LHC. We then close the paper with some concluding remarks in Section \ref{sec:discConc}.

\section{LHC data and unfolding} \label{sec:physics}

\subsection{Experimental data at the LHC} \label{sec:expData}

The Large Hadron Collider is a 27~km long circular proton-proton collider located in an underground tunnel at CERN in Geneva, Switzerland. 
With proton-proton collisions of up to 8 TeV\footnote{The electron volt, $\mathrm{ev}$, is the customary unit of energy used in particle physics, $1\ \mathrm{ev} \approx 1.6 \cdot 10^{-19}\ \mathrm{J}$.} center-of-mass energy, the LHC is the world's most powerful particle accelerator. The protons are accelerated in bunches of billions of particles and bunches moving in opposite directions are led to collide at the center of four gigantic particle detectors called ALICE, ATLAS, CMS and LHCb. In the current experimental configuration, these bunches collide every 50 ns at the heart of the detectors resulting in some 20 million collision events per second in each detector out of which the few hundred most interesting ones are stored for further analysis.

Out of the four detectors, ATLAS and CMS are multipurpose experiments capable of performing a large variety of physics analyses ranging from the discovery of the Higgs boson to precision studies of quantum chromodynamics. The other two detectors, ALICE and LHCb specialize in studies of lead-ion collisions and $b$-hadrons, respectively. In what follows, we focus on describing the data collection and analysis in the CMS experiment, which is also the source of the data of our unfolding demonstration in Section \ref{sec:Zboson}, but similar principles also apply to ATLAS and, to some extent, to other high energy physics experiments.

The CMS experiment \citep{CMS2008}, an acronym for Compact Muon Solenoid, is situated in an underground cavern along the LHC ring near the village of Cessy, France. The detector, weighing a total of 12\:500~tons, has a cylindrical shape with a diameter of 14.6~m and a length of 21.6~m. The construction, operation and data analysis of the experiment is conducted by an international collaboration of over 4000 scientists, engineers and technicians. When two protons collide at the center of CMS, their energy is transformed into matter in the form of new particles. A small fraction of these particles are exotic, short-lived particles, such as the Higgs boson or the top quark, which are at the center of the scientific interest of the high energy physics community. Such particles decay almost instantly into more familiar, stable particles, such as electrons, muons and photons. Using various subdetectors, the energies and trajectories of these particles are recorded in order to study the properties and interactions of the exotic particles created in the collision.

\begin{figure}[t]
%\vspace{6pc}
\centering
\includegraphics[trim = 0cm 0cm 0cm 0cm, clip=true, width=12.5cm]{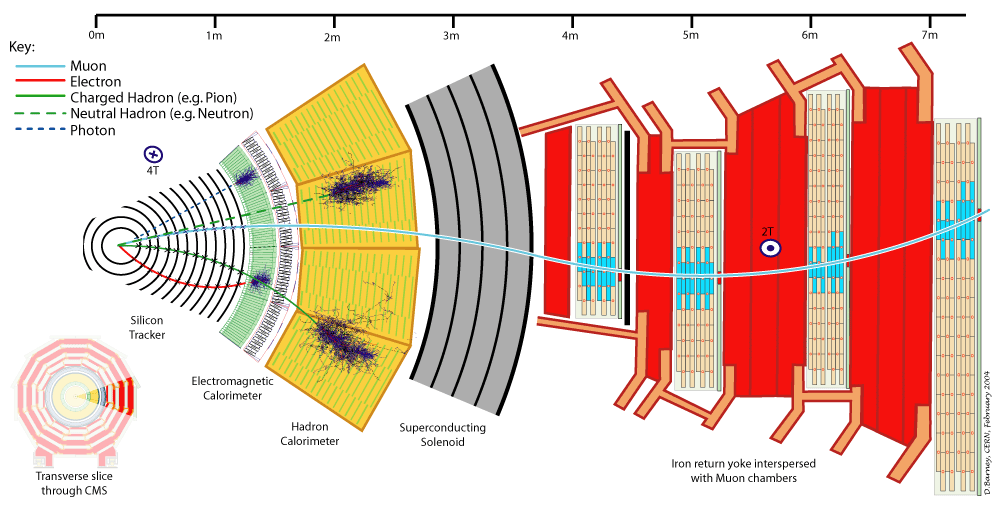}
\caption[]{Illustration of the detection of particles at the CMS experiment \citep{Barney2004}. Each type of a particle leaves its characteristic trace in the various subdetectors of the experiment. This enables identification of different particles as well as the measurement of their energies and trajectories. Copyright: CERN, for the benefit of the CMS Collaboration.}
\label{fig:CMS_slice}
\end{figure}

The layout of the CMS detector is illustrated in Figure \ref{fig:CMS_slice}. The detector is immersed in a 3.8~T magnetic field created using a superconducting solenoid magnet. This magnetic field bends the trajectory of any charged particle traversing the detector, and since the higher the momentum of the particle, the less it bends, this enables the measurement of its momentum. CMS consists of three layers of subdetectors: the tracker, the calorimeters and the muon detectors. The innermost detector is the silicon tracker, which consists of an inner layer of pixel detectors and an outer layer of microstrip detectors. When a charged particle passes through these semiconducting detectors, it leaves behind electron-hole pairs and hence creates an electric signal. These signals are combined into a particle track using a Kalman filter in order to reconstruct the trajectory of the particle.

The next layer of detectors are the calorimeters, which are devices for measuring the energies of particles. The CMS calorimeter system is divided into an electromagnetic calorimeter (ECAL) and a hadron calorimeter (HCAL). Both of these devices are based on the same general principle: they are made of extremely dense materials with the aim of stopping the particles passing through. In the process, a portion of the energy of these particles is converted into light in a scintillating material and the amount of light, which depends on the energy of the incoming particle, is measured using photodetectors inside the calorimeters. The ECAL measures the energy of particles that interact mostly via the electromagnetic interaction, in other words, electrons, positrons and photons. The HCAL, on the other hand, measures the energies of hadrons, i.e., particles composed of quarks. These include, e.g., protons, neutrons and pions. The HCAL is also instrumental in measuring the energies of jets, i.e., collimated streams of hadrons produced by quarks and gluons, and in detecting the so-called missing transverse energy, an energy imbalance caused by non-interacting particles, such as neutrinos, escaping the detector.

The outermost layer of the CMS detector consists of muon detectors, whose task is to identify and measure the momenta of muons. Accurate detection of muons was of central importance in the design of CMS since muons provide a clean signature for many exciting physics processes. This is because there is a very low probability for other particles, with the exception of non-interacting neutrinos, to penetrate through the CMS calorimeter system. For example, the four-muon decay channel played an important role in the discovery of the Higgs boson at CMS \citep{CMS2012Higgs}.

The information of all CMS subdetectors is combined \citep{CMS2009} to identify the stable particles, i.e., muons, electrons, positrons, photons and various types of hadrons, produced in each collision event, see Figure \ref{fig:CMS_slice}. For example, a muon will leave a track in both the silicon tracker and the muon chamber, while a photon produces a signal in the ECAL without an associated track in the tracker. The information of these individual particles is then used to reconstruct higher-level physics objects, such as jets and missing transverse energy.

\subsection{The role of unfolding in LHC data analysis}

The need for unfolding arises because any quantity measured by the detectors outlined above is corrupted by stochastic noise. For example, let $E$ be the energy of an electron hitting the CMS ECAL. Then the measured value of the energy follows to a good approximation the Gaussian distribution $\mathcal{N}(E,\sigma^2(E))$ where the variance satisfies \citep{CMS2008}
\begin{equation}
\left( \frac{\sigma(E)}{E} \right)^2 = \left( \frac{S}{\sqrt{E}} \right)^2 + \left( \frac{N}{E} \right)^2 + C^2, \label{eq:ECAL_res}
\end{equation}
where $S$, $N$ and $C$ are fixed constants. The noise is not necessarily additive. Furthermore, for more sophisticated measurements, such as the ones combining information from several subdetectors or more than one particle, the distribution of the response is not usually available in a closed form. Indeed, most analyses rely on detector simulations to determine the response of their physical quantity of interest.

It should be pointed out that not all LHC physics analyses directly rely on unfolding. The common factor between the examples listed in Section~\ref{sec:intro} is that these are {\em measurement\/} analyses and not {\em discovery\/} analyses meaning that these are analyses studying in detail the properties of some already known phenomenon. In such a case, the experimental interest often lies in the detailed physical shape of some distribution for which nonparametric unfolding is the appropriate tool to use, while discovery analyses almost exclusively use parametric models in the smeared space. The importance of unfolding for discovery of new physics lies in the fact that many unfolded results are either directly or indirectly used as inputs to discovery analyses. An example of this are {\em parton distribution functions\/} \citep{Forte2013} which quantify the internal structure of a proton. These functions are estimated via fits to unfolded spectra and are then used to derive theory predictions in various discovery analyses. They, for example, played an important role in the recent discovery of the Higgs boson \citep{ATLAS2012Higgs,CMS2012Higgs} and are vital in further searches of new physics, such as dark matter and extra dimensions \citep{CMS2012DM_ExtraDim}.% as well as the composite structure of quarks \citep{CMS2013CI}.

The need to unfold the measurements usually arises for the purposes of:

\begin{itemize}
\item \textbf{Comparison of experiments with different responses:} The only direct way of comparing the spectra measured in two different experiments, such as ATLAS and CMS, is to compare the unfolded measurements.
\item \textbf{Input to a subsequent analysis:} Certain tasks, such as the estimation of parton distributions functions and fine-tuning of Monte Carlo event generators, typically require unfolded input spectra. %The results directly influence a wide range of physics analyses including the important discovery analyses.
\item \textbf{Comparison with future theories:} When unfolded spectra are published, theorists can directly use them to compare with any new theoretical predictions which might not have existed at the time of the original measurement. This use case is sometimes considered controversial since alternatively one could publish the response of the detector and the theorists could use it to smear their new predictions.
\item \textbf{Exploratory data analysis:} The unfolded spectrum could reveal hidden structure in the data which is not considered in any of the existing theoretical predictions.
\end{itemize}

According to the CERN Document Server (\url{https://cds.cern.ch/}), the CMS experiment published in 2012 a total of 103 papers out of which 16 made direct use of unfolding and many more indirectly relied on unfolded results. Unfolding was most often used in studies of quantum chromodynamics (4 papers), forward physics (4) and properties of the top quark (3). Most of these results relied on the questionable bin-by-bin heuristic (8), while the EM algorithm (3) and various forms of penalization (6) were also used. We expect similar statistics to also hold for the other LHC experiments.

\section{Problem formulation} \label{sec:formulation}

In most situations in high energy physics, the data generation mechanism can be modelled as a {\em Poisson point process} (see, e.g. \citet{Reiss1993}). Let $E$ be a compact interval on $\mathbb{R}$, $f$ a non-negative function in $L^2(E)$ and $M$ a discrete random measure on $E$. Then $M$ is a Poisson point process on state space $E$ with intensity function $f$ if and only~if:
\begin{enumerate}
\item $M(B) \sim \mathrm{Poisson}(\lambda(B))$ with $\lambda(B) = \ownint{B}{}{f(s)}{s}$ for every Borel set $B \subset E$;
\item $M(B_1),\ldots,M(B_n)$ are independent for pairwise disjoint Borel sets $B_i \subset E, \, i=1,\ldots,n$.
\end{enumerate}
In other words, the number of points $M(B)$ observed in the set $B \subset E$ is Poisson distributed with mean $\ownint{B}{}{f(s)}{s}$ and the number of points in disjoint sets are independent random variables.

For the problem at hand, the Poisson process $M$ represents the true, particle-level observables generated in the proton-proton collisions. The \linebreak smeared, detector-level observables are represented by another Poisson process $N$. The process $N$ is assumed to have a state space $F$, which is a compact interval on $\mathbb{R}$, and a non-negative intensity function $g \in L^2(F)$. The intensities of the two processes are related by a bounded linear operator $K:L^2(E) \rightarrow L^2(F)$ so that $g = Kf$. In what follows, we assume $K$ to be a Fredholm integral operator, that is,
\begin{equation}
g(t) = (Kf)(t) = \ownint{E}{}{k(t,s)f(s)}{s}, \label{eq:FredIntEq}
\end{equation}
where the kernel $k \in L^2(F \times E)$ is assumed to be known. The unfolding problem is then to estimate the true intensity $f$ given a single observation of the smeared Poisson process $N$. % Should the spaces be restricted to non-negative L^2 functions only? Positive definite kernel?

This Poisson inverse problem \citep{Antoniadis2006, Reiss1993} is ill-posed in the sense that in virtually all practical cases the pseudoinverse $\pinv{K}$ of the forward operator $K$ is an unbounded --and hence discontinuous-- linear operator \citep{Engl2000}. This means that the na\"ive approach of first estimating $g$ using, for example, a kernel density estimate $\hat{g}$ and then estimating $f$ using $\hat{f} = \pinv{K}\hat{g}$ is unstable with respect to fluctuations of $\hat{g}$. The resulting na\"ive estimator has a huge variance which typically exhibits itself as large, unnatural oscillations in the estimates.

To better understand the physical meaning of the kernel $k$, let us consider the unfolding problem at the point level. Denoting by $X_i$ the true observables, the Poisson point process $M$ can be written as
\begin{equation}
M = \sum_{i=1}^\tau \delta_{X_i},
\end{equation}
where $\delta_{X_i}$ is the Dirac measure at $X_i \in E$ , the variables $\tau,X_1,X_2,\ldots$ are independent random variables such that $\tau \sim \mathrm{Poisson}(\lambda(E))$, and the $X_i$ are identically distributed with the probability density $f(\cdot)/\lambda(E)$, where $\lambda(E) = \ownint{E}{}{f(s)}{s}$.

When the particles corresponding to $X_i$ traverse the detector, the first thing that can happen is that they might not be observed at all due to the limited efficiency and acceptance of the device. Mathematically, this corresponds to {\em thinning} of the Poisson process. Let $Z_i \in \{0,1\}$ be an indicator variable showing whether the point $X_i$ is observed ($Z_i = 1$) or not ($Z_i = 0$). We assume that $\tau,(X_1,Z_1),(X_2,Z_2),\ldots$ are independent and that the pairs $(X_i,Z_i)$ are identically distributed. Then the thinned true process is given~by
\begin{equation}
M^* = \sum_{i=1}^\tau Z_i \delta_{X_i} = \sum_{i=1}^\xi \delta_{X_i^*},
\end{equation}
where $\xi = \sum_{i=1}^\tau Z_i$ and the $X_i^*$ are the true points with $Z_i = 1$. Denoting $\varepsilon(s) = P(Z_i = 1 | X_i = s)$, one can show that $M^*$ is a Poisson point process with intensity function $f^*(s) = \varepsilon(s) f(s)$.

For each observed point $X_i^* \in E$, the detector measures a noisy value $Y_i \in F$. We assume that the smeared observations $Y_i$ are i.i.d. with probability density
\begin{equation}
p(Y_i = t) = \ownint{E}{}{p(Y_i = t| X_i^* = s) p(X_i^* = s)}{s}.
\end{equation}
From this, it follows that the smeared observations $Y_i$ constitute a Poisson point process
\begin{equation}
N = \sum_{i=1}^\xi \delta_{Y_i}
\end{equation}
whose intensity function $g$ is given by
\begin{equation}
g(t) = \ownint{E}{}{p(Y_i=t|X_i^*=s)\varepsilon(s)f(s)}{s}.
\end{equation}
We hence identify that the kernel $k$ of Equation \eqref{eq:FredIntEq} is given by
\begin{equation}
k(t,s) = p(Y_i=t|X_i^*=s)\varepsilon(s).
\end{equation}
%Note that
%\begin{equation}
%\ownint{F}{}{k(t,s)}{t} = \varepsilon(s) \in [0,1],
%\end{equation}
%that is, given $s \in E$, we do not assume $k$ to integrate to unity over its first argument.

\section{Empirical Bayes unfolding} \label{sec:EBU}

\subsection{Outline of the proposed methodology} \label{sec:outline}

In this section, we propose a novel combination of statistical methods for solving the high energy physics unfolding problem formalized in Section \ref{sec:formulation}. The proposed methodology is based on the following four key ingredients:

\begin{enumerate}
 \item Discretization of the unknown particle-level intensity using a B-spline basis expansion, that is,
\begin{equation}
 f(s) = \sum_{j=1}^p \beta_j B_j(s), \quad s \in E,
\end{equation}
where $B_j(s),\,j=1,\ldots,p$, are the B-spline basis functions.
 \item Bayesian posterior mean estimation of the unknown basis coefficients $\bm{\beta} = \tp{\begin{bmatrix} \beta_1, \ldots, \beta_p \end{bmatrix}}$ using a single-component Metropolis--Hastings sampler.
 \item Empirical Bayes estimation of the scale $\delta$ of the regularizing smoothness prior $p(\bm{\beta}|\delta)$ using a Monte Carlo expectation-maximization algorithm.
 \item Frequentist uncertainty quantification and bias correction using the parametric bootstrap.
\end{enumerate}

This methodology enables a principled solution of the unfolding problem, including the choice of the regularization strength and uncertainty quantification, without having to resort to heuristics or approximations. We explain below each of these steps in detail and argue why this particular choice of techniques provides a natural framework for solving the problem at hand.

\subsection{Discretization of the problem} \label{sec:discretization}

Poisson inverse problems are almost exclusively studied in a form where for the observable process $N$ and the unobservable process $M$ are discretized. Usually the first step is to discretize the observable process using a histogram. In many applications this has to be done due to the discrete nature of the detector. In our case, the observations are, at least in principle, continuous, but we still carry out the discretization due to computational reasons. Indeed, in many analyses, there can be millions of observed collision events and treating each of these individually would not be computationally feasible.

In order to discretize the smeared process $N$, let $\{F_i\}_{i=1}^n$ be a partition of the smeared space $F$ into $n$ ordered intervals and let $y_i$ denote the number of points falling on interval $F_i$, that is, $y_i = N(F_i),\,i=1,\ldots,n$. This can be seen as recording the observed points in a histogram with bin contents \linebreak $\bm{y} = \tp{\begin{bmatrix} y_1,\ldots,y_n \end{bmatrix}}$ and is indeed the form of discretization most often employed in HEP. This discretization is convenient since it now follows from $N$ being a Poisson process that the $y_i$ are independent and Poisson distributed with means
\begin{equation}
 \mu_i = \ownint{F_i}{}{g(t)}{t} = \ownint{F_i}{}{\ownint{E}{}{k(t,s)f(s)}{s}}{t}, \quad i = 1,\ldots,n. \label{eq:smearedMean}
\end{equation}

In the true space $E$, there is no need to settle only for histograms. Instead, we consider a basis expansion of the true intensity $f$, that is,
\begin{equation}
 f(s) = \sum_{j=1}^p \beta_j \phi_j(s), \quad s \in E,
\end{equation}
where $\{\phi_j\}_{i=1}^p$ is a sufficiently large dictionary of basis functions. %Choosing $\phi_j(s) = \mathbf{1}_{E_j}(s)$, where $\{E_j\}_{j=1}^p$ is a partition of the true space $E$ with $p$ ordered intervals, we get a histogram representation for $f$, but with a more suitable choice of the basis, it is possible construct a smooth representation for $f$ with fewer free parameters than in the histogram representation.

Substituting the basis expansion of $f$ into Equation \eqref{eq:smearedMean}, we find that the means $\mu_i$ are given by
\begin{equation}
 \mu_i = \sum_{j=1}^p \left( \ownint{F_i}{}{\ownint{E}{}{k(t,s)\phi_j(s)}{s}}{t} \right) \beta_j = \sum_{j=1}^p K_{i,j} \beta_j,
\end{equation}
where we have denoted
\begin{equation}
 K_{i,j} = \ownint{F_i}{}{\ownint{E}{}{k(t,s)\phi_j(s)}{s}}{t}, \quad i=1,\ldots,n, \quad j=1,\ldots,p. \label{eq:Kij}
\end{equation}
Consequently, unfolding reduces to estimating $\bm{\beta}$ in the Poisson regression problem
\begin{equation}
 \bm{y}|\bm{\beta} \sim \mathrm{Poisson}(\bm{K}\bm{\beta}) \label{eq:PoissonRegressionProblem}
\end{equation}
for an ill-conditioned matrix $\bm{K} = (K_{i,j})$.

Since spectra in high energy physics are typically smooth functions, splines \citep{deBoor2001, Schumaker2007, Wahba1990} provide a particularly attractive way of representing the unknown intensity $f$. Let $\min E = s_0 < s_1 < s_2 < \cdots < s_L < s_{L+1} = \max E$ be a sequence of knots in the true space $E$. Then an order-$m$ spline with knots $s_i,\,i=0,\ldots,L+1$, is a piecewise polynomial whose restriction to each interval $[s_i,s_{i+1}),\,i=0,\ldots,L$, is an order-$m$ polynomial (i.e., a polynomial of degree $m-1$) and which has $m-2$ continuous derivatives at each interior knot $s_i,\,i=1,\ldots,L$. An order-$m$ spline with $L$ interior knots has $p=L+m$ degrees of freedom. In this work, we use exclusively order-4 cubic splines which consist of third degree polynomials and are twice continuously differentiable. Note also that an order-1 spline gives us the histogram representation of $f$.

There exist various bases $\{\phi_j\}_{j=1}^p$ for expressing splines of arbitrary order. We use B-splines $B_j,\,j=1,\ldots,p$, that is, spline basis functions of minimal local support, because of their numerical stability and conceptual simplicity. \citet{OSullivan1986,OSullivan1988} was among the first authors to use regularized B-spline estimators in statistical applications, with the approach later popularized by \citet{Eilers1996}. In the HEP unfolding literature, penalized maximum likelihood estimation with B-splines goes back to the work of \cite{Blobel1985} and recent contributions using similar methodology include \citet{Dembinski2011} and \citet{Milke2013}. We use the {\sc Matlab} Curve Fitting Toolbox to efficiently evaluate and perform basic operations on B-splines. These algorithms rely on the recursive use of lower-order B-spline basis functions, for details, see \citet{deBoor2001}.

The non-negativity of the intensity function $f$ is enforced by constraining $\bm{\beta}$ to be in $\mathbb{R}_+^p = \{ x \in \mathbb{R}^p : x_i \geq 0,\, i=1,\ldots,p \}$. This restricts $f$ to be non-negative since each of the B-spline basis functions $B_j,\,j=1,\ldots,p$, is non-negative.

\subsection{Bayesian estimation of the spline coefficients} \label{sec:estimation}

% Due to the ill-posedness of the unfolding problem, we need regularize the problem by limiting the effective degrees of freedom of the unknown intensity $f$. With the B-spline basis expansion, one way of doing this would be to limit the number of knots so that $p \ll n$, but such fits would potentially lose fine structure in the data and we would be confronted with the very difficult problem of choosing the number and locations of the knots. Instead, we take the approach often favored in contemporary statistics where we overparametrize the problem with a large number of knots and then regularize with a smoothness penalty. For B-splines such approach was first proposed by \citet{OSullivan1986, OSullivan1988} and later popularized by \citet{Eilers1996}.

In contrast to most work on unfolding, we take a Bayesian approach to estimation of the spline coefficients $\bm{\beta}$. That is, we estimate $\bm{\beta}$ using the Bayesian posterior
\begin{equation}
p(\bm{\beta}|\bm{y},\delta) = \frac{p(\bm{y}|\bm{\beta})p(\bm{\beta}|\delta)}{p(\bm{y}|\delta)} = \frac{p(\bm{y}|\bm{\beta})p(\bm{\beta}|\delta)}{\ownint{\mathbb{R}_+^p}{}{p(\bm{y}|\bm{\beta}')p(\bm{\beta}'|\delta)}{\bm{\beta}'}}, \quad \bm{\beta} \in \mathbb{R}_+^p, \label{eq:BayesRule}
\end{equation}
where the likelihood is given by the Poisson regression model \eqref{eq:PoissonRegressionProblem},
\begin{equation}
p(\bm{y}|\bm{\beta}) = \prod_{i=1}^n \frac{( \sum_{j=1}^p K_{i,j}\beta_j )^{y_i}}{y_i!} e^{-\sum_{j=1}^p K_{i,j} \beta_j}, \quad \bm{\beta} \in \mathbb{R}_+^p.
\end{equation}
The prior $p(\bm{\beta}|\delta)$, which regularizes the otherwise ill-posed problem, depends on a scale parameter $\delta$, which is analogous to the regularization parameter in the classical inverse problems literature.

We decided to use the Bayesian approach for two reasons. First, it provides a natural interpretation for the regularization via the prior density $p(\bm{\beta}|\delta)$, which should be chosen in such a way that most of its probability mass lies in physically plausible regions of the parameter space $\mathbb{R}_+^p$. Second, the Bayesian framework enables a principled, data-driven way of choosing the regularization strength $\delta$ using empirical Bayes estimation as explained below in Section~\ref{sec:empiricalBayes}.

In order to regularize the problem, we consider the truncated Gaussian smoothness prior
\begin{align}
p(\bm{\beta}|\delta) &\propto \exp \left( -\delta \|f''\|_2^2 \right) \\
&= \exp \left( -\delta \ownint{E}{}{\left\{f''(s)\right\}^2}{s} \right) \\
&= \exp \left( -\delta \tp{\bm{\beta}} \bm{\Omega} \bm{\beta} \right), \quad \bm{\beta} \in \mathbb{R}_+^p, \quad \delta > 0, \label{eq:smoothnessPrior}
\end{align}
where the elements of the $p \times p$ matrix $\bm{\Omega}$ are given by $\Omega_{i,j} = \ownint{E}{}{B_i''(s) B_j''(s)}{s}$. The interpretation of this prior is that the total curvature of $f$, characterized by $\|f''\|_2^2$, should be small. In other words, $f$ should be a relatively smooth function, which is true for most intensities encountered in high energy physics. The strength of the regularization is controlled by the hyperparameter $\delta$ --- the larger the value of $\delta$, the smoother $f$ is required to be.

The prior as defined by Equation \eqref{eq:smoothnessPrior} does not enforce any boundary conditions for the unknown intensity $f$. In this case, the matrix $\bm{\Omega}$ has rank $p-2$ and hence the prior is potentially improper (this depends on the orientation of the null space of $\bm{\Omega}$). Although the posterior would still be a proper probability density, the rank deficiency of $\bm{\Omega}$ is undesirable since the empirical Bayes approach requires a proper prior distribution. Furthermore, without any boundary constraints, the unfolded intensity has an unnecessarily large variance near the boundaries.

To address these issues, we use {\em Aristotelian boundary conditions} \citep{Calvetti2006}, where the idea is to condition the smoothness penalty on the boundary values $f(s_0)$ and $f(s_{L+1})$ and then place additional hyperpriors for these values. Since $f(s_0) = \beta_1 B_1(s_0)$ and $f(s_{L+1}) = \beta_p B_p(s_{L+1})$, we can equivalently condition on $(\beta_1,\beta_p)$. As a result, the prior model becomes
\begin{equation}
p(\bm{\beta}|\delta) = p(\beta_2,\ldots,\beta_{p-1}|\beta_1,\beta_p,\delta)p(\beta_1|\delta)p(\beta_p|\delta), \quad \bm{\beta} \in \mathbb{R}_+^p,
\end{equation}
where $p(\beta_2,\ldots,\beta_{p-1}|\beta_1,\beta_p,\delta) \propto \exp \left( -\delta \tp{\bm{\beta}} \bm{\Omega} \bm{\beta} \right)$. We model the boundaries using once again truncated Gaussians:
\begin{align}
&p(\beta_1|\delta) \propto \exp \left(-\delta \gamma_\mathrm{L} \beta_1^2 \right), \quad \beta_1 \geq 0, \\
&p(\beta_p|\delta) \propto \exp \left(-\delta \gamma_\mathrm{R} \beta_p^2 \right), \quad \beta_p \geq 0,
\end{align}
where $\gamma_\mathrm{L}, \gamma_\mathrm{R} > 0$ are fixed constants. The full prior can then be written as
\begin{equation}
p(\bm{\beta}|\delta) \propto \exp \left( -\delta \tp{\bm{\beta}} \bm{\Omega}_\mathrm{A} \bm{\beta} \right), \quad \bm{\beta} \in \mathbb{R}_+^p, \label{eq:AristotelianPrior}
\end{equation}
where the elements of the $p \times p$ matrix $\bm{\Omega}_\mathrm{A}$ are given by
\begin{equation}
\Omega_{\mathrm{A},i,j} = \begin{cases} \Omega_{i,j} + \gamma_\mathrm{L}, \quad &\text{if } i=j=1, \\ \Omega_{i,j} + \gamma_\mathrm{R}, \quad &\text{if } i=j=p, \\ \Omega_{i,j}, \quad &\text{otherwise}. \end{cases}
\end{equation}
The augmented matrix $\bm{\Omega}_\mathrm{A}$ is positive definite and hence Equation \eqref{eq:AristotelianPrior} defines a proper probability density. %Can this be shown analytically?

Once the hyperparameter $\delta$ has been estimated using empirical Bayes (see Section \ref{sec:empiricalBayes}), we plug its estimate $\hat{\delta}$ into Bayes' rule \eqref{eq:BayesRule} to obtain the empirical Bayes posterior $p(\bm{\beta}|\bm{y},\hat{\delta})$. We then use the mean of this posterior as a point estimator $\bm{\hat{\beta}}$ of the spline coefficients $\bm{\beta}$, that is, $\bm{\hat{\beta}} = \E \!\big(\bm{\beta}|\bm{y},\hat{\delta}\big)$, yielding the estimator $\hat{f}(s) = \sum_{j=1}^p \hat{\beta}_j B_j(s)$ of the unknown intensity $f$.
%Instead of the posterior mean, we could have also opted for maximum a posteriori (MAP) estimation. However, we decided to use the posterior mean for the following reasons:
%\begin{enumerate}
%\item The posterior mean is the decision rule which minimizes the Bayes risk, or equivalently the posterior expected loss, under the $l_2$ loss function (see, e.g., Section 4.4.2 of \cite{Berger1985}).
%%There is a decision theoretic justification for using the posterior mean. Namely, it is the Bayes rule under the $l_2$ loss function (see, e.g., Section 3.2 of \citet{Young2005}).
%\item In emission tomography, the posterior mean has been empirically found to give better or, in the worst case, similar finite-sample image reconstructions when compared to MAP estimation \citep{Geman1987,Green1996}.
%\item The posterior mean can be easily obtained from the same Markov chain Monte Carlo sampler used to find the empirical Bayes estimate of $\delta$, while finding the MAP estimate would require implementation of an additional numerical procedure.
%%\item Heuristically, the posterior mean is sensitive to posterior probability mass, while the MAP estimate is sensitive to posterior probability density.
%\end{enumerate}

Of course, in practice, the posterior $p(\bm{\beta}|\bm{y},\delta)$ is not available in a closed form because of the intractable integral in the denominator of Bayes' rule~\eqref{eq:BayesRule}. Hence, we need to resort to Markov chain Monte Carlo (MCMC) \citep{Robert2004} sampling from the posterior and the posterior mean is then computed as the empirical mean of the Monte Carlo sample. Unfortunately, the most elementary MCMC samplers are not well-suited for solving the problem at hand: Gibbs sampling is not computationally tractable since the full posterior conditionals do not belong to any of the standard families of probability distributions and the Metropolis--Hastings sampler with multivariate proposals is difficult to implement since the posterior can have very different scales for different components of $\bm{\beta}$.

To be able to efficiently sample from the posterior, we adopt the single-component Metropolis--Hastings sampler (also known as the Metropolis-within-Gibbs sampler) proposed by \citet{Saquib1998}. Denoting $\bm{\beta}_{-k} = \tp{\begin{bmatrix} \beta_1, \ldots, \beta_{k-1}, \beta_{k+1}, \ldots, \beta_p \end{bmatrix}}$, the basic idea of the sampler is to approximate the full posterior conditionals $p(\beta_k|\bm{\beta}_{-k},\bm{y},\delta)$ of the Gibbs sampler using a more tractable density \citep{Gilks1996MCMC,Gilks1996FullCond}. One then samples from this approximate full conditional and performs a Metropolis--Hastings acceptance step to correct for the approximation error. In our case, we take a second-order Taylor expansion of the non-quadratic part of the log full conditional resulting in a Gaussian approximation of the full conditional. When the mean of this Gaussian is non-negative, we sample from its truncation to the non-negative real line, and if the mean is negative, we replace the Gaussian tail by an exponential distribution. Further details on the MCMC sampler can be found in Section~III.C of \citet{Saquib1998}.

During each iteration of the Monte Carlo expectation-maximization algorithm used in the empirical Bayes estimation of $\delta$ (see Section \ref{sec:empiricalBayes}), we verify the convergence and mixing of the MCMC sampler by monitoring the acceptance rates of the Metropolis--Hastings proposals and the autocorrelation times $\kappa_j,\,j=1,\ldots,p$, of the Markov chain. The latter measure how often the sampler on average produces an independent observation from the posterior and is estimated using Geyer's initial convex sequence estimator (ICSE) \citep{Geyer1992} computed using the {\sc R} package \texttt{mcmc} \citep{Geyer2013}. The autocorrelation times $\kappa_j$ enable us to define the effective sample sizes $\mathrm{ESS}_j = S/\kappa_j,\,j=1,\ldots,p$, where $S$ is the size of the MCMC sample. $\mathrm{ESS}_j$ measures the effective number of independent observations obtained for the $j$th component of the Markov chain \citep[p. 99]{Kass1998}. For the MCMC iteration producing the final point estimate~$\bm{\hat{\beta}}$, we also monitor the trace plots, histograms, estimated autocorrelation functions and cumulative means of each component $\beta_j,\,j=1,\ldots,p$, of the Markov chain.

\subsection{Empirical Bayes selection of the regularization strength} \label{sec:empiricalBayes}

The Bayesian approach to solving inverse problems is particularly attractive since it admits selection of the regularization strength $\delta$ using marginal maximum likelihood estimation. For a comprehensive introduction to this and related empirical Bayes methods, see, e.g., Chapter 5 of \citet{Carlin2009}. The main idea in empirical Bayes is to regard the marginal distribution $p(\bm{y}|\delta)$ appearing in the denominator of Bayes' rule \eqref{eq:BayesRule} as a parametric model for the data $\bm{y}$ and then use standard frequentist point estimation techniques to estimate the hyperparameter $\delta$. 

The {\em marginal maximum likelihood estimator} (MMLE) of the hyperparameter $\delta$ is defined as the maximizer of $p(\bm{y}|\delta)$ with respect to $\delta$. That is, we estimate $\delta$ using
\begin{equation}
 \hat{\delta} = \argmax_{\delta > 0} p(\bm{y}|\delta) = \argmax_{\delta > 0} \ownint{\mathbb{R}_+^p}{}{p(\bm{y}|\bm{\beta})p(\bm{\beta}|\delta)}{\bm{\beta}}. \label{eq:defMMLE}
\end{equation}
Computing the MMLE is non-trivial since we cannot evaluate the high-dimensional integral in \eqref{eq:defMMLE} either in a closed form or using standard numerical integration methods. Monte Carlo integration, where one samples $\{\bm{\beta}^{(s)}\}_{s=1}^{S}$ from the prior $p(\bm{\beta}|\delta)$ and then approximates
\begin{equation}
 p(\bm{y}|\delta) \approx \frac{1}{S} \sum_{s=1}^{S} p(\bm{y}|\bm{\beta}^{(s)}), \quad \bm{\beta}^{(s)} \overset{\mathrm{i.i.d.}}{\sim} p(\bm{\beta}|\delta), \label{eq:naiveMCIntegration}
\end{equation}
is also out of question. This is because, in the high-dimensional parameter space, most of the $\bm{\beta}^{(s)}$'s fall on regions where the likelihood $p(\bm{y}|\bm{\beta}^{(s)})$ is numerically zero. Hence we would need an enormous sample size $S$ to get even a rough idea of the marginal likelihood $p(\bm{y}|\delta)$.

Luckily, it is possible to circumvent these issues by using the expectation-maximization (EM) algorithm \citep{Dempster1977,McLachlan2008} to find the MMLE. In the context of Poisson inverse problems, this approach was originally proposed by \citet{Geman1985, Geman1987} for tomographic image reconstruction and later studied and extended by \cite{Saquib1998}, but has received little attention since then. When applied to the unfolding problem, the standard EM prescription reads as follows. Let $(\bm{y},\bm{\beta})$ be the complete data, in which case the complete-data log-likelihood is given by
\begin{equation}
 l(\delta; \bm{y},\bm{\beta}) = \log p(\bm{y},\bm{\beta}|\delta) = \log p(\bm{y}|\bm{\beta}) + \log p(\bm{\beta}|\delta),
\end{equation}
where we have used $p(\bm{y},\bm{\beta}|\delta) = p(\bm{y}|\bm{\beta})p(\bm{\beta}|\delta)$. In the E-step of the algorithm, one computes the expectation of the complete-data log-likelihood over the unknown spline coefficients $\bm{\beta}$ conditional on the observations $\bm{y}$ and the current hyperparameter $\delta^{(t)}$:
\begin{align}
 Q(\delta;\delta^{(t)}) &= \E \!\big(l(\delta; \bm{y},\bm{\beta}) \big| \bm{y},\delta^{(t)} \big) \\
&= \E \!\big(\log p(\bm{y},\bm{\beta}|\delta) \big| \bm{y},\delta^{(t)} \big) \\
&= \E \!\big(\log p(\bm{\beta}|\delta) \big| \bm{y},\delta^{(t)}\big) + \mathrm{const},
\end{align}
where the constant does not depend on $\delta$. In the subsequent M-step, one maximizes the expected complete-data log-likelihood $Q(\delta;\delta^{(t)})$ with respect to the hyperparameter $\delta$. This maximizer is then used as the hyperparameter on the next step of the algorithm:
\begin{equation}
 \delta^{(t+1)} = \argmax_{\delta > 0} Q(\delta;\delta^{(t)}) = \argmax_{\delta > 0} \E \!\big(\log p(\bm{\beta}|\delta) \big| \bm{y},\delta^{(t)}\big). \label{eq:MStep}
\end{equation}
By Theorem 1 of \citet{Dempster1977}, each step of this iteration is guaranteed to increase the incomplete-data likelihood $p(\bm{y}|\delta)$, that is, $p(\bm{y}|\delta^{(t+1)}) \geq p(\bm{y}|\delta^{(t)}),\,t=0,1,2,\ldots$ With this construction, the incomplete-data likelihood conveniently coincides with the marginal likelihood and hence the EM algorithm enables us find the MMLE %\footnote{In more rigorous terms, the monotonicity of the marginal likelihood sequence $\{p(\bm{y}|\delta^{(t)})\}$ does not imply the convergence of the hyperparameter sequence $\{\delta^{(t)}\}$ to the MMLE without additional assumptions and studies which are beyond the scope of this paper.}
$\hat{\delta}$ of the hyperparameter $\delta$.

The expectation in Equation \eqref{eq:MStep},
\begin{equation}
 \E \!\big(\log p(\bm{\beta}|\delta) \big| \bm{y},\delta^{(t)}\big) = \ownint{\mathbb{R}_+^p}{}{p(\bm{\beta}|\bm{y},\delta^{(t)})\log p(\bm{\beta}|\delta)}{\bm{\beta}}, \label{eq:expectationIntegral}
\end{equation}
again involves an intractable integral, but can be computed using Monte Carlo integration. We simply need to sample $\{\bm{\beta}^{(s)}\}_{s=1}^S$ from the posterior $p(\bm{\beta}|\bm{y},\delta^{(t)})$ and then replace the expectation by its Monte Carlo approximation:
\begin{equation}
 \E \!\big(\log p(\bm{\beta}|\delta) \big| \bm{y},\delta^{(t)}\big) \approx \frac{1}{S} \sum_{s=1}^S \log p(\bm{\beta}^{(s)}|\delta), \quad \bm{\beta}^{(s)} \sim p(\bm{\beta}|\bm{y},\delta^{(t)}). \label{eq:MCIntegration}
\end{equation}
The posterior sample can be obtained using the single-component Metropolis--Hastings sampler described in Section \ref{sec:estimation}.
%The convergence of the empirical mean to the true conditional expectation is then guaranteed by the ergodicity of the MCMC sampler.
The resulting variant of the EM~algorithm is called a {\em Monte Carlo expectation-maximization {\em (MCEM)} algorithm} \citep{Wei1990}. Due to the inevitable Monte Carlo error on each E-step, the MCEM algorithm loses the monotonicity property of the standard EM algorithm and theoretical analysis of its convergence becomes involved. However, in certain special cases, the iteration has been shown to eventually reach an arbitrarily small neighborhood of the maximizer with a high probability \citep{Chan1995}.

To summarize, the MCEM algorithm for finding the MMLE of the hyperparameter $\delta$ iterates between the following two steps:
\begin{description}
 \item[E-step:]Sample $\bm{\beta}^{(1)},\ldots,\bm{\beta}^{(S)}$ from the posterior $p(\bm{\beta}|\bm{y},\delta^{(t)})$ and compute
\begin{equation}
 \widetilde{Q}(\delta;\delta^{(t)}) = \frac{1}{S} \sum_{s=1}^S \log p(\bm{\beta}^{(s)}|\delta). \label{eq:QTilde}
\end{equation}
 \item[M-step:] Set $\delta^{(t+1)} = \argmax_{\delta > 0} \widetilde{Q}(\delta;\delta^{(t)})$.
\end{description}

This MCEM algorithm has a rather intuitive interpretation. First, on the E-step, we use the current iterate $\delta^{(t)}$ to produce a sample of $\bm{\beta}$'s from the posterior. Since this sample summarizes our current best understanding of $\bm{\beta}$, we then tune the prior by varying $\delta$ on the M-step to match this sample as well as possible, and the value of $\delta$ that matches the posterior sample the best will then become the next iterate~$\delta^{(t+1)}$.

One could also wonder why Monte Carlo integration works for the expectation of Equation \eqref{eq:expectationIntegral} while it did not work for directly computing the marginal likelihood $p(\bm{y}|\delta)$ in Equation \eqref{eq:naiveMCIntegration}. There are at least two reasons for this. First, in the MCEM algorithm, the $\bm{\beta}$'s are sampled from the posterior and hence most of them correspond to reasonable unfolded intensities. This means that they should also lie within the region where the bulk of the prior probability mass is located, thus making the sample mean in Equation \eqref{eq:MCIntegration} well-behaved. On the contrary, in Equation~\eqref{eq:naiveMCIntegration}, the sample is generated from the prior resulting mostly in intensities that do not match the data very well. Second, the sum in \eqref{eq:naiveMCIntegration} is over plain densities instead of log-densities as in Equation \eqref{eq:MCIntegration}. This makes the MCEM computations considerably more robust against small probability density function values.

When $p(\bm{\beta}|\delta)$ is given by the Aristotelian smoothness prior \eqref{eq:AristotelianPrior}, the \linebreak M-step of the MCEM algorithm is available in a closed form. Taking normalization into account, the prior density is given by
\begin{equation}
 p(\bm{\beta}|\delta) = C(\delta) \exp(-\delta \tp{\bm{\beta}} \bm{\Omega}_\mathrm{A} \bm{\beta}),
\end{equation}
where the normalization constant $C(\delta)$ depends on the hyperparameter $\delta$ and satisfies
\begin{equation}
 C(\delta) = \frac{\delta^{p/2}}{\ownint{\mathbb{R}_+^p}{}{\exp(-\tp{\bm{\beta}}\bm{\Omega}_\mathrm{A}\bm{\beta})}{\bm{\beta}}}.
\end{equation}
Hence
\begin{equation}
 \log p(\bm{\beta}|\delta) = \frac{p}{2} \log \delta - \delta \tp{\bm{\beta}} \bm{\Omega}_\mathrm{A} \bm{\beta} + \mathrm{const},
\end{equation}
where the constant does not depend on $\delta$. Plugging this into Equation \eqref{eq:QTilde}, we find that the maximizer on the M-step is given by
\begin{equation}
 \delta^{(t+1)} = \frac{1}{\frac{2}{pS} \sum_{s=1}^S \tp{(\bm{\beta}^{(s)})} \bm{\Omega}_\mathrm{A} \bm{\beta}^{(s)}}.
\end{equation}

The resulting iteration for finding the MMLE $\hat{\delta}$ is summarized in Algorithm~\ref{alg:MCEM}. The MCMC sampler is started from the empirical mean of the posterior sample of the previous iteration in order to facilitate the convergence of the Markov chain. In this work, we run the MCEM algorithm for a fixed number of steps $T$, but one could easily devise more elaborate stopping rules for the algorithm. Note, however, that the optimal choice of this stopping rule and the MCMC sample size $S$ are, to a large extent, open problems \citep{Booth1999}.

\begin{algorithm}
\caption{MCEM algorithm for finding the MMLE}
\label{alg:MCEM}
\begin{algorithmic}
\Require
\State $\bm{y}$ --- Observed data
\State $\delta^{(0)} > 0$ --- Initial guess 
\State $T$ --- Number of MCEM iterations
\State $S$ --- Size of the MCMC sample
\State $\bm{\beta}_\mathrm{init}$ --- Starting point for the MCMC sampler
\Ensure
\State $\hat{\delta}$ --- MMLE of the hyperparameter $\delta$
\State
\State Set $\bar{\bm{\beta}} = \bm{\beta}_\mathrm{init}$
\For{$t = 1$ \algorithmicto\ $T$} 
\State Sample $\bm{\beta}^{(1)},\bm{\beta}^{(2)},\ldots,\bm{\beta}^{(S)} \sim p(\bm{\beta}|\bm{y},\delta^{(t-1)})$ starting from $\bar{\bm{\beta}}$ using the single-component Metropolis--Hastings sampler of \cite{Saquib1998}
\State Set
\begin{equation*}
 \delta^{(t)} = \frac{1}{\frac{2}{pS} \sum_{s=1}^S \tp{(\bm{\beta}^{(s)})} \bm{\Omega}_\mathrm{A} \bm{\beta}^{(s)}}
\end{equation*}
\State Compute $\bar{\bm{\beta}} = \sum_{s=1}^S \bm{\beta}^{(s)}$
\EndFor
\State \Return $\hat{\delta} = \delta^{(T)}$
\end{algorithmic}
\end{algorithm}

\subsection{Uncertainty quantification and bias correction} \label{sec:UQ}

The final ingredient of our procedure is uncertainty quantification and bias correction of the estimated intensity $\hat{f}$. In contrast to most other applications of Poisson inverse problems, uncertainty quantification is of vital importance in our problem setting. It turns out that, because of our use of empirical Bayes, uncertainty quantification of $\hat{f}$ is not entirely straightforward. For example, credible intervals based on the empirical Bayes posterior $p(\bm{\beta}|\bm{y},\hat{\delta})$ lose their subjective Bayesian interpretation because of the data-driven choice of the hyperparameter $\delta$. Also, such intervals do not take into account uncertainty regarding the choice of $\delta$ and their frequentist properties are poorly understood.

There has been a fair amount of work on correcting the na\"{i}ve empirical Bayes confidence intervals (EBCI) obtained using the posterior $p(\bm{\beta}|\bm{y},\hat{\delta})$ to account for the uncertainty of $\hat{\delta}$ (see Section 5.4 of \citet{Carlin2009}), including the bootstrap technique of \citet{Laird1987}. This work, however, is aimed at achieving coverage with respect to the hierarchical sampling model $p(\bm{y},\bm{\beta}|\delta) = p(\bm{y}|\bm{\beta})p(\bm{\beta}|\delta)$, while in our case standard frequentist coverage with respect to $p(\bm{y}|\bm{\beta})$ would arguably be a more desirable goal. This is because in our case the prior $p(\bm{\beta}|\delta)$ is introduced simply to regularize the ill-posedness of the problem and does not take part in the actual physical process generating the data $\bm{y}$.

We propose quantifying the uncertainty of $\hat{f}$ using a parametric bootstrap technique which is distinct from that of \citet{Laird1987} by aiming for confidence intervals with standard frequentist coverage. The approaches we propose are similar to those of \citet{Cowling1996} but extend their results to the case of an indirectly observed Poisson point process. Our starting point is to regard the estimator $\bm{\hat{\beta}}$ as a frequentist point estimator of $\bm{\beta}$, that is, $\bm{\hat{\beta}} = \bm{\hat{\beta}}(\bm{y}) = \E \!\big(\bm{\beta}|\bm{y},\hat{\delta}(\bm{y}) \big)$. We then resample the data $\bm{y}$ and plug in the resampled observations $\bm{y}^*$ to obtain the resampled estimates $\bm{\hat{\beta}}^* = \bm{\hat{\beta}}(\bm{y}^*) = \E \!\big(\bm{\beta}|\bm{y}^*,\hat{\delta}(\bm{y}^*) \big)$.

\begin{figure}[t]
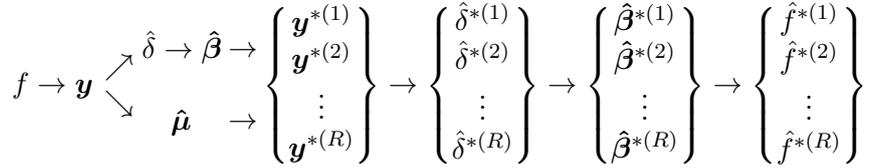

%\vspace{6pc}
\begin{equation*}
f \rightarrow \bm{y} \: \begin{matrix} \, \nearrow \\ \, \searrow \end{matrix} \begin{matrix} \; \hat{\delta} \rightarrow \bm{\hat{\beta}} \\ \\ \: \bm{\hat{\mu}} \end{matrix} \begin{matrix} \: \rightarrow \\ \\ \: \rightarrow \end{matrix} \begin{Bmatrix}\bm{y}^{*(1)} \\ \bm{y}^{*(2)} \\ \vdots \\ \bm{y}^{*(R)} \end{Bmatrix} \rightarrow \begin{Bmatrix} \hat{\delta}^{*(1)} \\ \hat{\delta}^{*(2)} \\ \vdots \\ \hat{\delta}^{*(R)} \end{Bmatrix} \rightarrow \begin{Bmatrix} \bm{\hat{\beta}}^{*(1)} \\ \bm{\hat{\beta}}^{*(2)} \\ \vdots \\ \bm{\hat{\beta}}^{*(R)} \end{Bmatrix} \rightarrow \begin{Bmatrix} \hat{f}^{*(1)} \\ \hat{f}^{*(2)} \\ \vdots \\ \hat{f}^{*(R)} \end{Bmatrix}
\end{equation*}
\caption[]{Illustration of the bootstrap procedure for generating a resample $\hat{f}^{*(r)},\,r=1,\ldots,R$, of unfolded intensities. Resampling can either be based on $\bm{\hat{\beta}}$ (Scheme 1) or $\bm{\hat{\mu}}$ (Scheme 2).}
\label{fig:bootstrapIllustration}
\end{figure}

In our problem setting, one can envisage several different resampling schemes to obtain the bootstrapped observations $\bm{y}^*$. In particular, we consider the following two parametric resampling procedures:
\begin{description}
 \item[Scheme 1:] Resample $\bm{y}^* \overset{\mathrm{i.i.d.}}{\sim} \mathrm{Poisson}(\bm{K}\bm{\hat{\beta}})$, where $\bm{\hat{\beta}} = \E \!\big(\bm{\beta}|\bm{y},\hat{\delta}(\bm{y})\big)$, our empirical Bayes point estimate of the spline coefficients $\bm{\beta}$.
 \item[Scheme 2:] Resample $\bm{y}^* \overset{\mathrm{i.i.d.}}{\sim} \mathrm{Poisson}(\bm{\hat{\mu}})$, where $\bm{\hat{\mu}} = \bm{y}$, the maximum likelihood estimate of the smeared means $\bm{\mu}$.
\end{description}
Of these, the former corresponds to Method 1 of \citet{Cowling1996} and the latter to their Method 2. Irrespective of the resampling method used, we rerun the MCEM algorithm for each $\bm{y}^*$ to find the bootstrapped hyperparameter $\hat{\delta}^* = \hat{\delta}(\bm{y}^*)$. By doing this, we are able to also take into account the uncertainty regarding the choice of the hyperparameter $\delta$. The resampled spline coefficients are then found as the mean of the bootstrapped posterior $\bm{\hat{\beta}}^* = \E \! \big(\bm{\beta}|\bm{y}^*,\hat{\delta}^*\big)$ resulting in the bootstrapped unfolded intensity $\hat{f}^*(s) = \sum_{j=1}^p \hat{\beta}^*_j B_j(s)$. This procedure is then repeated $R$ times to obtain a sample of bootstrapped intensities $\mathcal{F}^* = \{\hat{f}^{*(r)}\}_{r=1}^R$. The resulting bootstrap procedure is illustrated in Figure~\ref{fig:bootstrapIllustration}.

Various techniques have been proposed for constructing confidence bands for $f$ based on the bootstrap sample $\mathcal{F}^*$, see \citet{Efron1993} and \citet{Davison1997}. Letting $\hat{f}_\alpha^*(s)$ denote the $100 \cdot \alpha$th percentile of the sample $\mathcal{F}^*$ evaluated at~$s \in E$, we form pointwise confidence bands for $f$ using the following two standard techniques:
\begin{description}
 \item[Basic bootstrap interval:] For every $s \in E$, an approximate $1-2\alpha$ confidence interval for $f(s)$ is given by $[2\hat{f}(s) - \hat{f}_{1-\alpha}^*(s),2\hat{f}(s) - \hat{f}_{\alpha}^*(s)]$.
 \item[Percentile interval:] For every $s \in E$, an approximate $1-2\alpha$ confidence interval for $f(s)$ is given by $[\hat{f}_\alpha^*(s),\hat{f}_{1-\alpha}^*(s)]$.
\end{description}
%The underlying philosophy with the basic bootstrap interval is to use the bootstrap distribution $p(\hat{f}^*(s)|\hat{f}(s))$ in the Neyman construction of confidence intervals \citep{Neyman1937}, while the percentile intervals are formed after a normalizing transformation \citep{Efron1993}.

% The bootstrap sample $\mathcal{F}^*$ can be used to construct confidence bands for $f$ in various ways. The simplest approach is to form pointwise confidence bands using percentile intervals. That is, for every $s \in E$, $[\hat{f}_\alpha^*(s),\hat{f}_{1-\alpha}^*(s)]$ serves as an approximate $1-2\alpha$ confidence interval for $f(s)$, where $\hat{f}_\alpha^*(s)$ denotes the $100 \cdot \alpha$th percentile of the sample $\mathcal{F}^*$ evaluated at~$s$.

%In addition to pointwise confidence bands, we can also use the sample $\mathcal{F}^*$ to quantify the uncertainty of other functionals of $\hat{f}$. For example, if the unfolded intensity contains a resonance peak, we can form an approximate confidence interval for the location of the peak by considering the bootstrap confidence interval corresponding to the location of the appropriate mode of the bootstrapped intensities $\hat{f}^{*(r)},\,r=1,\ldots,R$.

Choosing between resampling schemes 1 and 2 and basic and percentile intervals is tricky since there exists no clear consensus on their relative merits and superiority \citep{Cowling1996, Davison1997, Efron1993}. Scheme 2 will produce bootstrapped estimates $\hat{f}^*$ which follow closely the actual sampling distribution of $\hat{f}$. As such, we found that $\E \! \big( \hat{f}^* | \bm{y} \big) \approx \hat{f}$ which invalidates the use of the bootstrap to recover the bias of $\hat{f}$. Furthermore, when scheme 2 is used, there is usually little difference between the basic intervals and the percentile intervals. Under scheme 1, on the other hand, $\hat{f}^*$ will follow the sampling distribution of $\hat{f}$ conditional on the observed value of the estimator, hence enabling the bootstrap to probe the bias of $\hat{f}$. If a large bias is present in $\hat{f}$ and scheme 1 is used, the percentile intervals will perform poorly as they will be ``upside down'', while the basic intervals will implicitly account for the bias. We thus recommend the combination of scheme 1 and basic intervals be used if $\hat{f}$ is suspected to be significantly biased, while for large sample sizes with small biases, the conceptually simpler combination of scheme 2 and percentile intervals can also be used. Of course, if sufficient computational resources are available, the best would be to construct both of these combinations and see if they agree. If they do, either can be used, but if there is a disagreement, then the combination of scheme 1 with basic intervals is likely to be more trustworthy due to its ability to (partially) account for the bias.

%Furthermore, in situations where significant bias is present, percentile intervals are known to perform poorly while basic intervals implicitly take the bias into account. For large sample sizes with little bias in $\hat{f}$, we recommend using the conceptually simplest combination of Scheme 2 with percentile intervals, but if $\hat{f}$ is suspected to be significantly biased, we recommend using basic intervals with Scheme 1.

In the case of significant bias, it is also possible to apply a bootstrap bias correction to the point estimate $\hat{f}$. The standard bootstrap estimate of the bias of $\hat{f}$ at $s \in E$ is $\widehat{\mathrm{bias}}^*\!\big(\hat{f}(s)\big) = \frac{1}{R} \sum_{r=1}^R \hat{f}^*(s) - \hat{f}(s)$ which gives rise to the the bias-corrected point estimate $\hat{f}_\mathrm{BC}(s) = \hat{f}(s) - \widehat{\mathrm{bias}}^*\!\big(\hat{f}(s)\big)$. Note that, given the discussion above, bias correction only makes sense when resampling scheme 1 is used.

We conclude this section by noting that although using the bootstrap is computationally intensive, the computational cost can be alleviated through the use of parallel computing. Indeed, the bootstrap procedure outlined above is fully parallelizable since no communication is required between the individual bootstrap replications. We used the {\sc Matlab} Parallel Computing Toolbox to parallelize all the bootstrap computations reported below and generally obtained a roughly three-fold speed-up of the computations on a quad-core desktop computer setup.
%Using the {\sc Matlab} Parallel Computing Toolbox, a speed-up of 1.8 with respect to serial execution was obtained on a dual-core laptop and a speed-up of 3.1 on a quad-core desktop computer.

% One should note that the confidence intervals formed this way do not take into account the inevitable bias of $\hat{f}$. In other words, these are strictly speaking confidence intervals for $\E\!\big(\hat{f}|f\big)$ and not for the true intensity $f$ itself. Unfortunately, bias correction in this setting is, to a large extent, a difficult, open problem. For further discussion on this issue, see Section \ref{sec:discConc}.

% \begin{algorithm}
% \caption{Uncertainty quantification using parametric bootstrap}
% \label{alg:bootstrap}
% \begin{algorithmic}
% \Require
% \State $\bm{y}$ --- Observed data
% \State $R$ --- Size of the bootstrap sample
% \Ensure
% \State $\hat{f}^{*(1)},\hat{f}^{*(2)},\ldots,\hat{f}^{*(R)}$ --- Bootstrap sample of unfolded intensities
% \State
% \For{$r = 1$ \algorithmicto\ $R$}
% \State Sample $\bm{y}^{*(r)} \sim \mathrm{Poisson}(\bm{\hat{\mu}})$ with $\bm{\hat{\mu}} = \bm{y}$
% \State Find $\hat{\delta}^{*(r)}$ using Algorithm \ref{alg:MCEM}
% \State Estimate $\bm{\hat{\beta}}^{*(r)} = \E \! \big(\bm{\beta}|\bm{y}^{*(r)},\hat{\delta}^{*(r)}\big)$
% \State Store $\hat{f}^{*(r)} = \sum_{j=1}^p \hat{\beta}^{*(r)} B_j$
% \EndFor
% \State \Return $\hat{f}^{*(1)},\hat{f}^{*(2)},\ldots,\hat{f}^{*(R)}$
% \end{algorithmic}
% \end{algorithm}

\section{Simulation studies} \label{sec:simulations}

\subsection{Experiment setup} \label{sec:simSetup}

We first demonstrate the empirical Bayes unfolding methodology using simulated data. The data were generated using a two-component Gaussian mixture model on top of a uniform background and smeared by convolving the particle-level intensity with a Gaussian density. Specifically, the true process $M$ had the intensity
\begin{equation}
 f(s) = \lambda_\mathrm{tot} \left\{ \pi_1 \mathcal{N}(s|-2,1) + \pi_2 \mathcal{N}(s|2,1) + \pi_3 \frac{1}{|E|} \right\}, \quad s \in E,
\end{equation}
where $\lambda_\mathrm{tot} = \E(\tau) = \ownint{E}{}{f(s)}{s} > 0$ is the expected number of true observations, $|E|$ denotes the Lebesgue measure of $E$ and the mixing proportions $\pi_i$ sum up to one and were set to $\pi_1 = 0.2$, $\pi_2 = 0.5$ and $\pi_3 = 0.3$. The true space $E$ and the smeared space $F$ were both taken to be the interval $[-7,7]$. The true points $X_i$ were smeared with additive Gaussian noise of zero mean and unit variance. Points smeared beyond the boundaries of $F$ were discarded from further analysis. With this setup, the smeared intensity is given by the convolution
\begin{equation}
 g(t) = (Kf)(t) = \ownint{E}{}{\mathcal{N}(t-s|0,1)f(s)}{s}, \quad t \in F.
\end{equation}
Note that this setup corresponds to the classically most difficult class of deconvolution problems since the Gaussian error has a supersmooth probability density \citep{Meister2009}.

The smeared space $F$ was discretized using $n=40$ histogram bins of uniform size, while the true space $E$ was discretized using order-4 B-splines with $L = 26$ uniformly placed interior knots resulting in $p = L + 4 = 30$ unknown basis coefficients. With these choices, the condition number of the smearing matrix $\bm{K}$ was $\mathrm{cond}(\bm{K}) \approx 2.6 \cdot 10^8$ indicating that the problem is severely ill-posed. The boundary hyperparameters were set to $\gamma_\mathrm{L} = \gamma_\mathrm{R} = 5$. All experiments reported in this paper were implemented in {\sc Matlab} and the computations were carried out on a desktop setup with a quad-core 2.7~GHz Intel Core i5 processor.

\subsection{Results} \label{sec:simResults}

We first consider a relatively easy large-sample problem where $\lambda_\mathrm{tot} = 20\:000$. The MCEM algorithm was started using the initial hyperparameter $\delta^{(0)} = 1 \cdot 10^{-5}$ and was run for 20 iterations. The MCMC sampler was started from the non-negative least-squares spline fit to the smeared data, i.e., $\bm{\beta}_\mathrm{init} = \min_{\bm{\beta} \geq 0} \|\tilde{\bm{K}}\bm{\beta} - \bm{y}\|_2^2$, where the elements of $\tilde{\bm{K}}$ are given by Equation \eqref{eq:Kij} with the smearing kernel $k(t,s) = \delta_0(t-s)$. This problem is significantly less ill-posed than the unfolding problem --- the condition number of $\tilde{\bm{K}}$ was only 25. For each EM iteration, the single-component Metropolis--Hastings algorithm was used to obtain 500 post-burn-in observations from the posterior. After convergence of the EM algorithm, the final point estimate $\bm{\hat{\beta}}$ was obtained using a sample size of 1\:000. The whole procedure was then repeated with $R=200$ bootstrap replications obtained using resampling scheme 2. Running the MCEM iteration once to find the point estimate $\bm{\hat{\beta}}$ took 3 minutes, while the running time of the whole algorithm was 3 h 36 min with the bootstrap computations parallelized on the four cores of the quad-core setup.

\begin{figure}[!hp]
\centering
%\vspace{6pc}
\subfigure{
\includegraphics[trim = 0cm 0cm 0cm 0cm, clip=true, width=11.5cm]{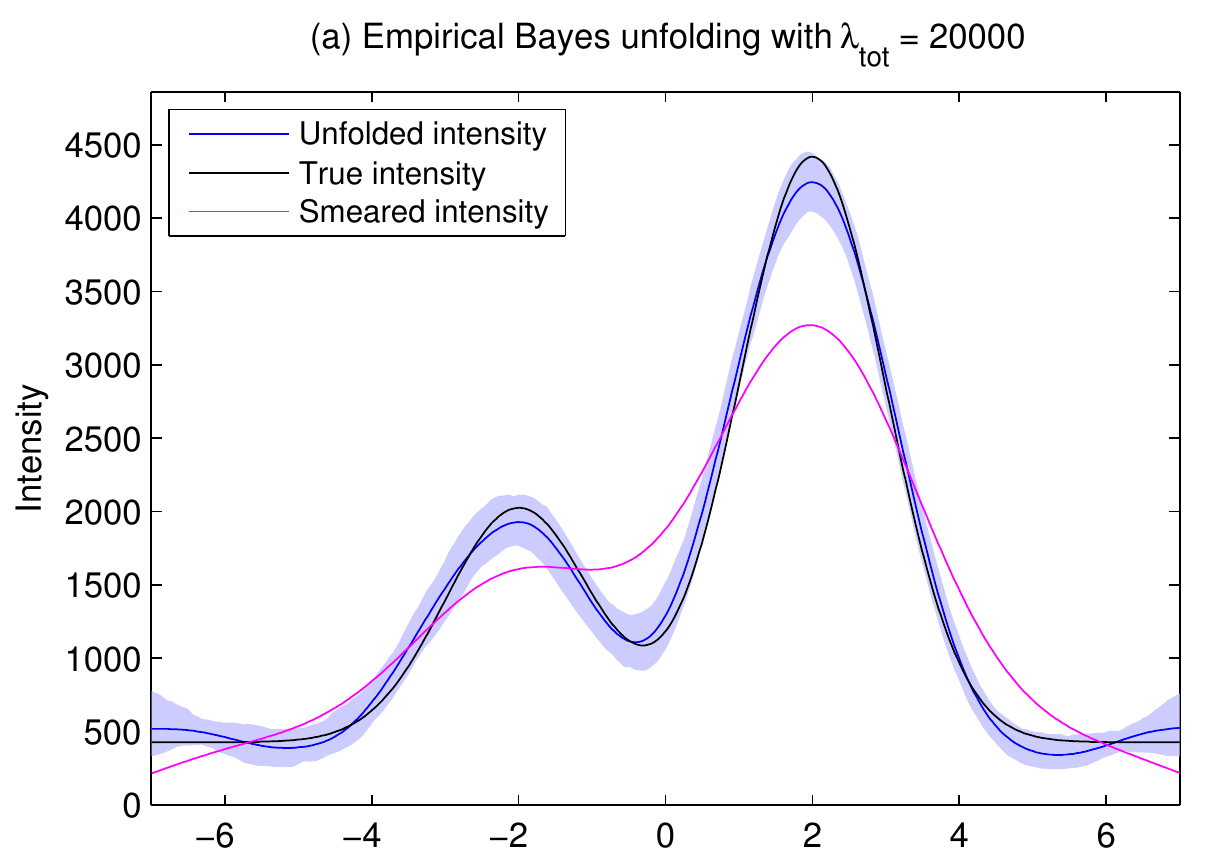}
\label{fig:gmm20000Reg}}
\subfigure{
\includegraphics[trim = 0cm 0cm 0cm 0cm, clip=true, width=11.5cm]{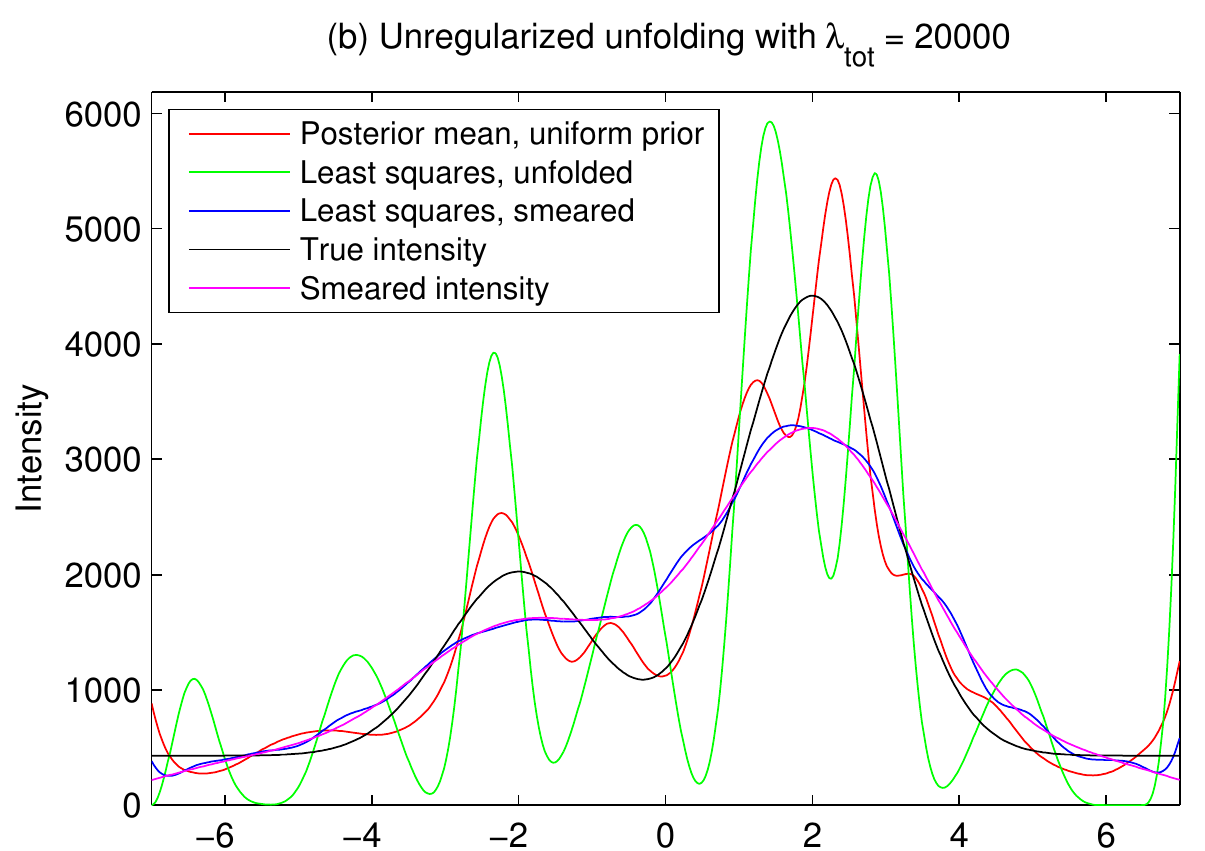}
\label{fig:gmm20000NoReg}}
\caption{Unfolding results for the Gaussian mixture model data with $\lambda_\mathrm{tot} = 20\:000$. Figure~(a) shows the unfolded intensity obtained using empirical Bayes unfolding along with 95\:\% pointwise percentile intervals. Figure~(b) illustrates the ill-posedness of the problem by showing the unfolded intensities obtained using non-negative least-squares estimation and posterior mean estimation with a uniform prior. Also shown is the least-squares fit to the smeared data.} 
\label{fig:gmm20000}
\end{figure}

\begin{figure}[!t]
\centering
%\vspace{6pc}
\subfigure{
\includegraphics[trim = 0cm 0cm 0cm 0cm, clip=true, width=6cm]{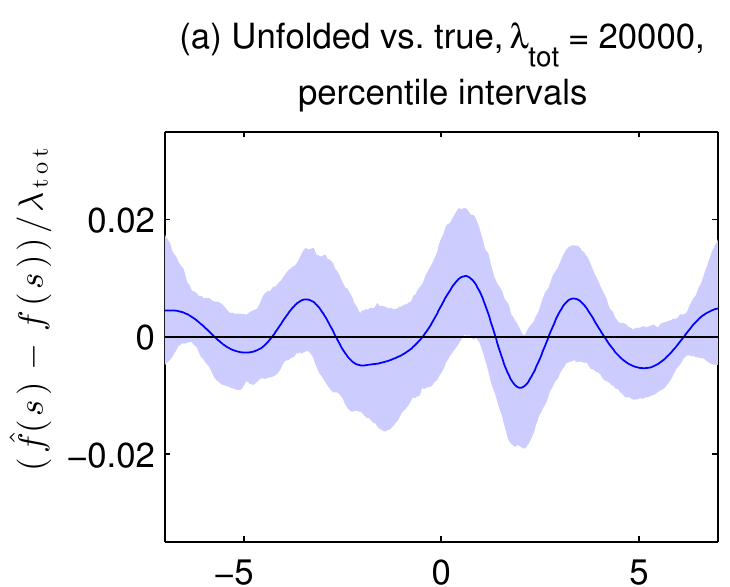}
\label{fig:gmm20000DiffBsPer}}
\subfigure{
\includegraphics[trim = 0cm 0cm 0cm 0cm, clip=true, width=6cm]{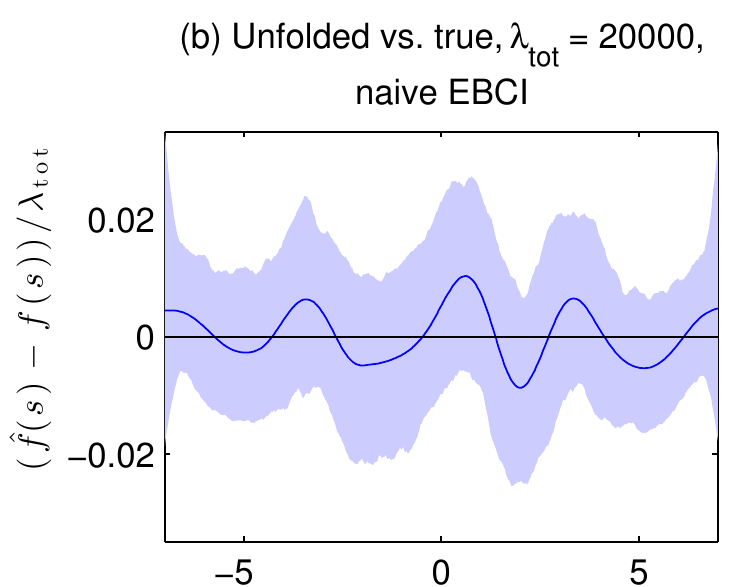}
\label{fig:gmm20000DiffEBCI}}
\caption{Difference between the unfolded intensity $\hat{f}$, obtained using empirical Bayes unfolding, and the true intensity $f$ normalized for the expected sample size $\lambda_\mathrm{tot} = 20\:000$. Figure~(a) shows the 95\:\% pointwise percentile intervals, while Figure~(b) shows the corresponding na\"{i}ve empirical Bayes confidence intervals.} 
\label{fig:gmm20000Diff}
\end{figure}

Figure \ref{fig:gmm20000Reg} shows the true intensity $f$, the smeared intensity $g$ and the unfolded intensity $\hat{f}$ with 95\:\% pointwise percentile intervals. The unfolded intensity beautifully captures the two peaks of the true intensity despite the severely corrupted observations. The only artifacts are the small wiggles on both tails of the intensity. Moreover, the percentile intervals cover the true $f$ for all values of $s \in E$ except for a short interval near $s = 0.7$. This is best seen in Figure \ref{fig:gmm20000DiffBsPer} where the confidence intervals are shown after subtraction of the true intensity $f$ and normalizing for the expected sample size $\lambda_\mathrm{tot}$. To enable comparison between the bootstrap and the empirical Bayes intervals, we have also plotted the na\"{i}ve empirical Bayes confidence intervals in Figure \ref{fig:gmm20000DiffEBCI}. These intervals seem to cover equally well, but are consistently longer then the percentile intervals which is likely to result in overcoverage.

\begin{figure}[ht]
\centering
%\vspace{6pc}
\subfigure{
\includegraphics[trim = 0cm 0cm 0cm 0cm, clip=true, width=6cm]{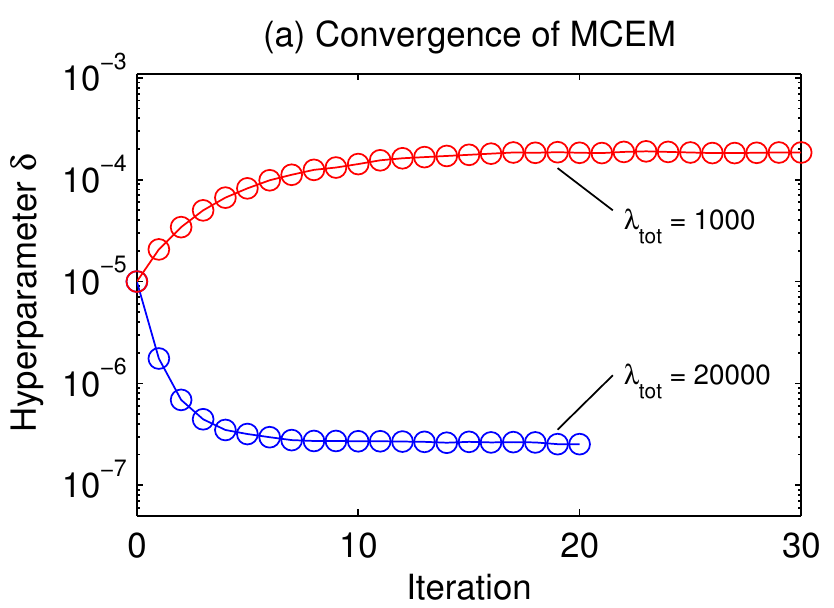}
\label{fig:gmm20000AlphaConv}}
\subfigure{
\includegraphics[trim = 0cm 0cm 0cm 0cm, clip=true, width=6cm]{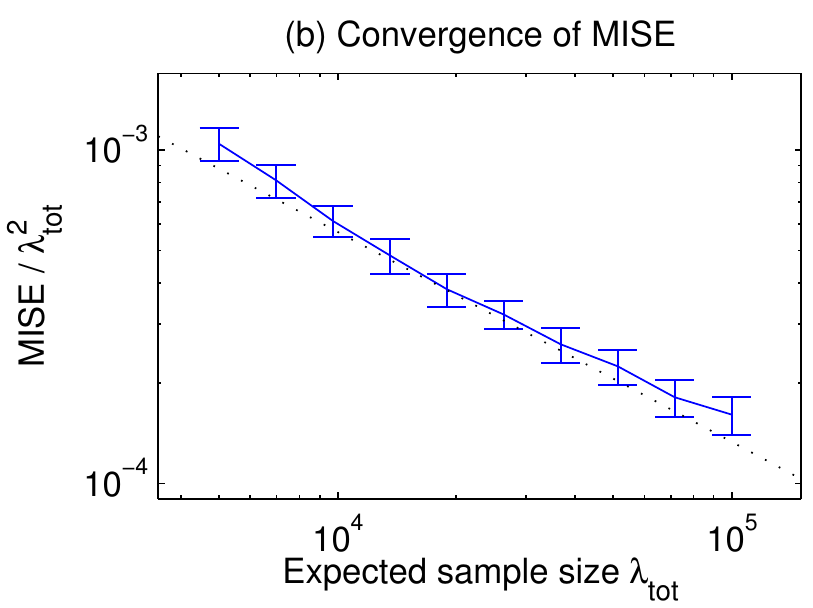}
\label{fig:gmmMISE}}
\caption{Convergence studies for empirical Bayes unfolding. Figure~(a) illustrates the convergence of the Monte Carlo EM algorithm and shows that the algorithm converges faster for larger sample sizes. Figure~(b) shows the convergence of the mean integrated squared error (MISE) as the expected sample size $\lambda_\mathrm{tot}$ grows. Note that convergence is only obtained for $\mathrm{MISE}/\lambda_\mathrm{tot}^2$. The error bars indicate approximate 95\:\% confidence intervals, %computed as $\pm 1.96 \cdot \widehat{\mathrm{SE}}$, where $\widehat{\mathrm{SE}}$ is the estimated standard error of the mean.
and the dotted straight line was added as a reference to illustrate that the convergence appears to be slightly slower than that given by a power law.}
\label{fig:gmmAlphaConvMISE}
\end{figure}

\begin{figure}[ht]
\centering
%\vspace{6pc}
\includegraphics[trim = 0cm 0cm 0cm 0cm, clip=true, width=11.8cm]{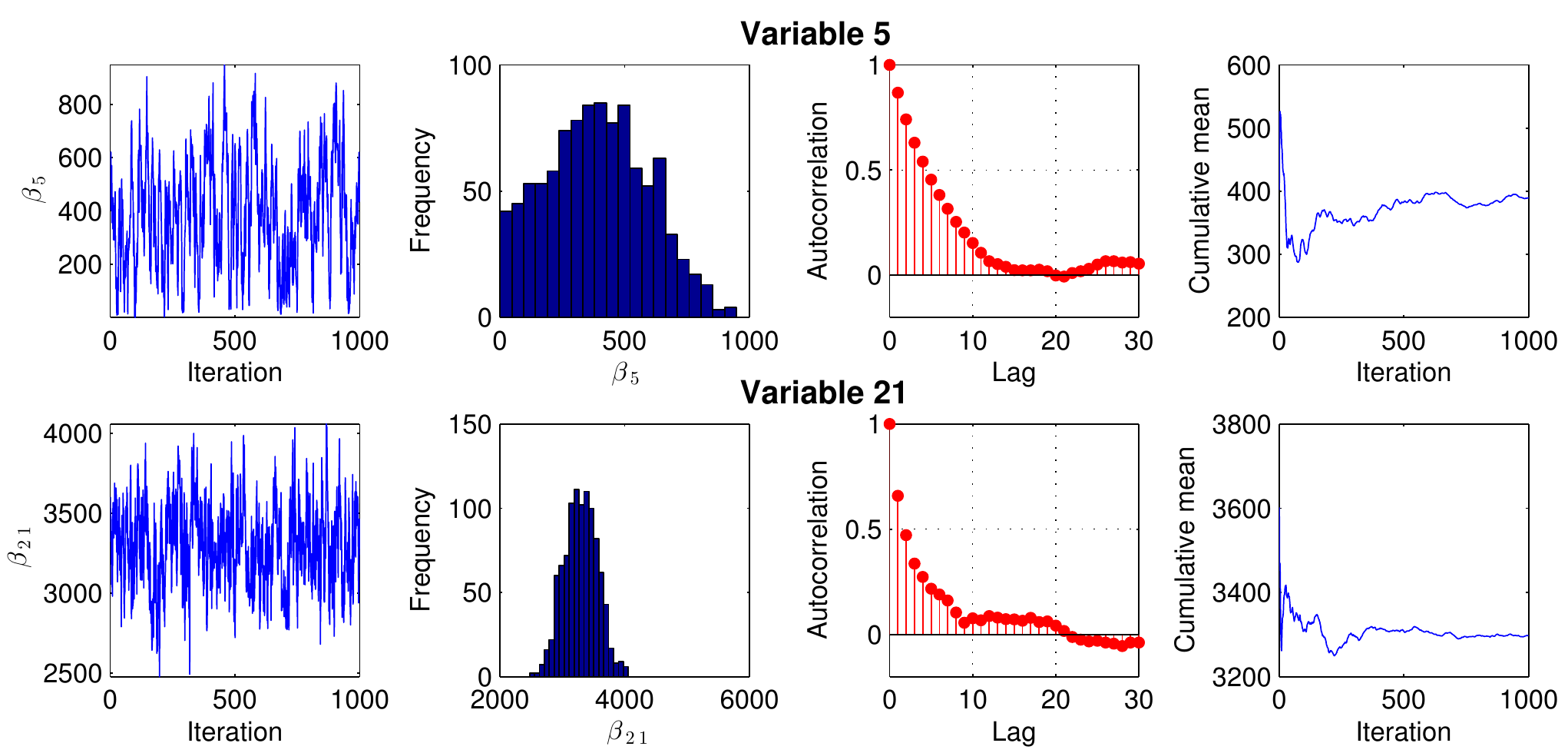}
\caption{Convergence and mixing diagnostics for the single-component Metropolis--Hastings sampler for variables $\beta_5$ and $\beta_{21}$: from left to right, the trace plots, histograms, estimated autocorrelation functions and cumulative means of the samples. For variable $\beta_5$ the acceptance rate was 97\:\%, the lag 1 autocorrelation 0.87 and the autocorrelation time 10.8. Hence the effective sample size for $\beta_5$ was 92.6. For $\beta_{21}$ the corresponding values were 99\:\%, 0.66 and 6.8 with the effective sample size 146.0.}
\label{fig:gmm20000MCMC}
\end{figure}

Figure \ref{fig:gmm20000AlphaConv} shows the convergence of the hyperparameter estimates of the MCEM algorithm. The algorithm reduced the regularization strength from the initial value to the final estimate $\hat{\delta} = 2.5 \cdot 10^{-7}$ and converged after approximately 10 iterations. During the iteration, the autocorrelation time of the MCMC sampler averaged over the components of $\bm{\beta}$ increased from 4.7 to 8.4 indicating that it was easier to sample from the more regularized posterior. A typical proposal acceptance rate was 98\:\%. %\footnote{As opposed to the standard multivariate Metropolis--Hastings algorithm, the single-component version of the algorithm tries to imitate the Gibbs sampler and hence the ideal acceptance rate would be 100\:\%.}
For the final MCMC run producing the point estimate $\bm{\hat{\beta}}$, a more careful performance analysis was made for each component of the sampler. Figure \ref{fig:gmm20000MCMC} shows the diagnostic plots for the components $\beta_5$ and $\beta_{21}$ after the removal of the burn-in. These plots indicate that the chain has converged and mixes reasonably well although the performance of the chain is typically slightly better in the interior of the space ($\beta_{21}$) than closer to the boundaries ($\beta_5$).

To illustrate the importance of regularization in solving this ill-posed problem, we also ran the MCMC sampler with the uniform prior $p(\bm{\beta}) \propto 1,\linebreak \bm{\beta} \in \mathbb{R}_+^p$. In the absence of regularization, the single-component Metropolis--Hastings algorithm had significant issues exploring the parameter space. Indeed, for a sample size 1\:000 (after a burn-in of 500 observations), the average autocorrelation time was 60.8 and the largest autocorrelation time 191.5 corresponding to only 5.2 effective observations from the posterior. This slow mixing was also apparent in the the trace plots and cumulative means of the chain. Unsurprisingly, the posterior mean computed based on this sample exhibits many undesired oscillations as seen in Figure \ref{fig:gmm20000NoReg}. The figure also depicts the non-negative least-squares solutions corresponding to the design matrices $\bm{K}$ and $\tilde{\bm{K}}$. The latter was used as the starting point $\bm{\beta}_\mathrm{init}$ of the MCMC iteration and is relatively well-behaved, but once one tries to undo the smearing, the solution falls apart.

In order to consider a more difficult test case, we repeated the experiment with the expected sample size $\lambda_\mathrm{tot} = 1\:000$. In this case, we found that the MCEM algorithm converged more slowly and that the hyperparameter estimates $\delta^{(t)}$ exhibited larger Monte Carlo variation. We hence increased the number of MCEM iterations to 30 and sampled 1\:000 observations from the posterior on each EM iteration. In addition, we used bootstrap resampling scheme 1 which enables us to probe the bias of the estimator. Otherwise the parameters of the experiment were the same as above. With these changes, obtaining the point estimate $\bm{\hat{\beta}}$ took 9 minutes and the full running time was 9 h 56 min.

\begin{figure}[t]
%\vspace{6pc}
\centering
\includegraphics[trim = 0cm 0cm 0cm 0cm, clip=true, width=11.5cm]{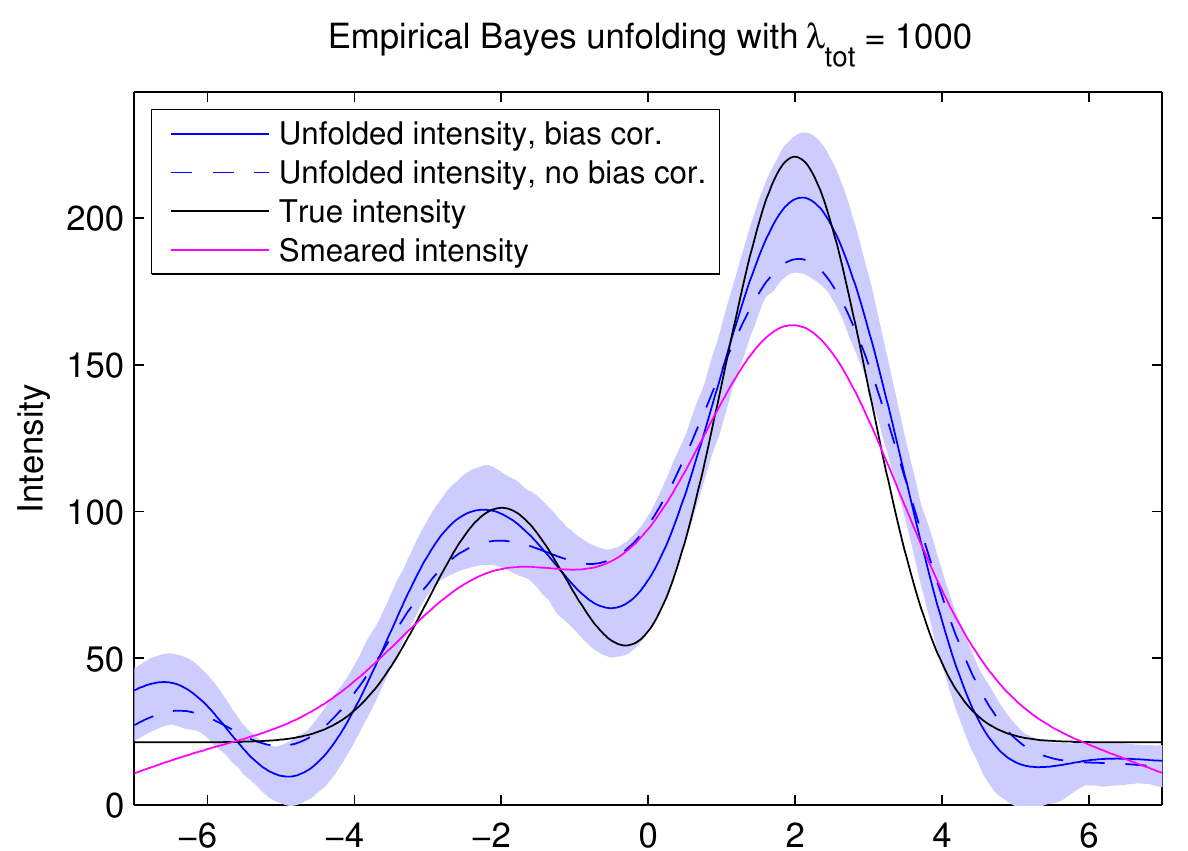}
\caption{Unfolding results for the Gaussian mixture model data with $\lambda_\mathrm{tot} = 1\:000$. The unfolded intensity obtained using empirical Bayes unfolding (dashed blue curve) is biased and hence bootstrap bias correction is applied (solid blue curve). The confidence band consists of 95\:\% pointwise basic bootstrap intervals.}
\label{fig:gmm1000Reg}
\end{figure}

\begin{figure}[!ht]
\centering
%\vspace{6pc}
\subfigure{
\includegraphics[trim = 0cm 0cm 0cm 0cm, clip=true, width=6cm]{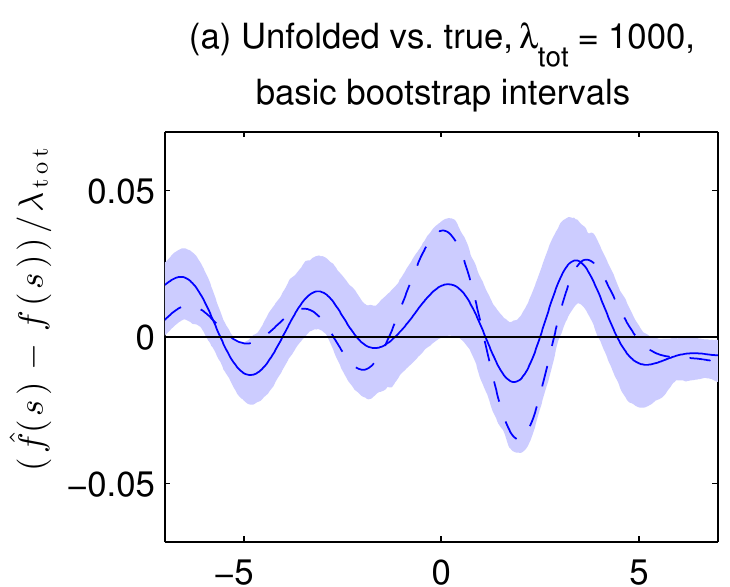}
\label{fig:gmm1000DiffBsBasic}}
\subfigure{
\includegraphics[trim = 0cm 0cm 0cm 0cm, clip=true, width=6cm]{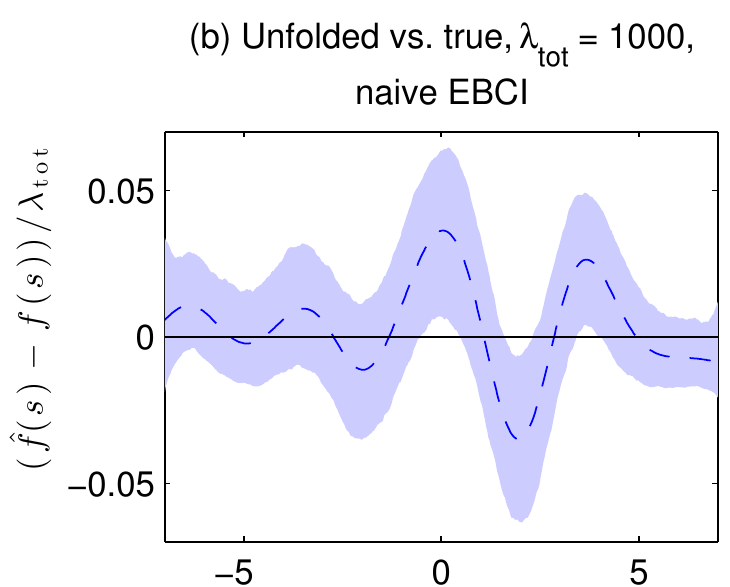}
\label{fig:gmm1000DiffEBCI}}
\caption{Difference between the unfolded intensity $\hat{f}$ and the true intensity $f$ normalized for the expected sample size $\lambda_\mathrm{tot} = 1\:000$. Figure~(a) shows the 95\:\% pointwise basic bootstrap intervals, while Figure~(b) shows the corresponding na\"{i}ve empirical Bayes confidence intervals. Both figures include the original point estimate $\hat{f}$ (dashed curve), and Figure~(a) also shows the bias-corrected estimate $\hat{f}_\mathrm{BC}$ (solid curve).} 
\label{fig:gmm1000Diff}
\end{figure}

Figure \ref{fig:gmm20000AlphaConv} illustrates that the MCEM iteration increased the regularization strength to $\hat{\delta} = 1.8 \cdot 10^{-4}$ and converged after approximately 20 iterations. During the iteration, the mean autocorrelation time increased from roughly 3.5 to 4.6 indicating that in this case it was slightly more difficult to sample from the more regularized posterior. The diagnostic plots for the final sampling did not indicate any problems with the convergence and mixing of the sampler. The resulting unfolded intensity $\hat{f}$ is represented by the dashed curve in Figure \ref{fig:gmm1000Reg}. The estimate is clearly biased near the peaks and the trough of the true intensity, but this can be mitigated with bootstrap bias correction. The bias-corrected estimate $\hat{f}_\mathrm{BC}$, shown as the solid curve, captures the shape of the true intensity significantly better than the original estimate, but, as always, this reduction in the bias comes at the cost of increased variance visible particularly near the boundaries of the space. The figure also shows the 95\:\% pointwise basic bootstrap intervals which seem to cover the true intensity reasonably well, albeit potentially at the price of some slight undercoverage\footnote{Note that due to the strong correlation between $\hat{f}(s_1)$ and $\hat{f}(s_2)$, when $s_1$ and $s_2$ are close to each other, one cannot draw conclusions regarding the coverage of the bootstrap intervals by simply looking at Figure \ref{fig:gmm1000DiffBsBasic}. Instead, one would have to repeat the whole inference procedure for several independent observations of $\bm{y}$ which would require enormous amounts of computing time.} (as suggested by Figure \ref{fig:gmm1000DiffBsBasic} where the estimates are plotted after subtracting the true intensity $f$ and normalizing for the expected sample size $\lambda_\mathrm{tot}$). Figure \ref{fig:gmm1000DiffEBCI} shows also the corresponding na\"{i}ve empirical Bayes confidence intervals. These intervals are longer than the basic bootstrap intervals, but as explained in Section \ref{sec:UQ}, their statistical interpretation is unclear. Note also between Figures \ref{fig:gmm20000Diff} and \ref{fig:gmm1000Diff} the improvement in the point estimate $\hat{f}$ and the reduction in the length of the confidence intervals when moving from the expected sample size $\lambda_\mathrm{tot} = 1\:000$ to $\lambda_\mathrm{tot} = 20\:000$.

To further study how empirical Bayes unfolding behaves as a function of the sample size, we repeated our first experimental setup on a logarithmic grid of expected sample sizes from $\lambda_\mathrm{tot} = 5\:000$ up to $\lambda_\mathrm{tot} = 100\:000$. For each sample size, we unfolded 100 independent smeared observations $\bm{y}$ and estimated the mean integrated squared error (MISE) of $\hat{f}$ as the sample mean of the integrated squared errors $\mathrm{ISE} = \ownint{E}{}{\,(\hat{f}(s)-f(s))^2}{s}$. As $\lambda_\mathrm{tot} \rightarrow \infty$, we expect the MISE to diverge, but $\mathrm{MISE}/\lambda_\mathrm{tot}^2$ should converge to zero, and this is indeed what we observe in Figure \ref{fig:gmmMISE}. On a log-log scale, the MISE estimates appear to slightly deviate from a straight line indicating that the convergence speed is likely to be close to a power law but slightly slower.

%\vspace{1cm}

\section{Unfolding of the $Z$ boson invariant mass spectrum} \label{sec:Zboson}

\subsection{Description of the data}

In this section, we illustrate empirical Bayes unfolding using real data from the CMS experiment at the Large Hadron Collider. In particular, we unfold the $Z$ boson invariant mass spectrum published in \citet{CMS2013ECAL}. The $Z$ boson, which is produced in copious quantities at the LHC, is a mediator of the weak interaction. The particle is very short-lived and decays almost instantly into other elementary particles. The decay mode considered here is the decay of a $Z$ boson into into a positron and an electron,\linebreak $Z \rightarrow e^+ e^-$. The original purpose of these data was to calibrate and measure the resolution of the CMS electromagnetic calorimeter but they also serve as an excellent testbed for unfolding since the true intensity of this spectrum is known with remarkable precision from previous experiments.

The electron and the positron produced in the decay of the $Z$ boson are first detected in the CMS silicon tracker after which their energies~$E_i,\linebreak i=1,2$, are measured by stopping the particles at the ECAL, see Section~ \ref{sec:expData}. From this information, one can compute the \emph{invariant mass} $W$ of the electron-positron system defined by the equation
\begin{equation}
W^2 = (E_1 + E_2)^2 - \|\bm{p}_1 + \bm{p}_2\|_2^2,
\end{equation}
where $\bm{p}_i,\,i=1,2$, are the momenta of the two particles and the equation is written using the natural units where the speed of light $c=1$. Since \linebreak $\|\bm{p}_i\|_2^2 = E_i^2 - m_e^2$, where $m_e$ is the rest mass of the electron, one can reconstruct the invariant mass $W$ using only the ECAL energy deposits $E_i$ and the opening angle between the two tracks in the silicon tracker.

The invariant mass $W$ is preserved in particle decays. Furthermore, it is invariant under Lorentz transformations and has therefore the same value in every frame of reference. This means that the invariant mass of the $Z$~boson, which is simply its rest mass $m$, is equal to the invariant mass of the electron-positron system, $W = m$. It follows that measurement of the invariant mass spectrum of the electron-positron pair enables us to measure the mass spectrum of the $Z$ boson itself.

Due to the time-energy uncertainty principle, the $Z$ boson does not have a unique rest mass $m$. Instead, the mass follows the Cauchy distribution, also known in particle physics as the \emph{Breit--Wigner distribution}, whose density is given by
\begin{equation}
p(m) = \frac{1}{2\pi} \frac{\Gamma}{(m-m_Z)^2 + \frac{\Gamma^2}{4}},
\end{equation}
where $m_Z = 91.1876\ \mathrm{GeV}$ is the mode of the distribution (often simply called \emph{the} mass of the $Z$ boson) and $\Gamma = 2.4952\ \mathrm{GeV}$ is the full width of the distribution at half maximum \citep{Beringer2012}. Since the contribution of background processes to the electron-positron channel near the $Z$ peak is negligible \citep{CMS2013ECAL}, the underlying true intensity $f(m)$ is proportional to $p(m)$.

The dominant source of smearing in measuring the $Z$ boson invariant mass~$m$ is the measurement of the energy deposits $E_i$ in the ECAL. The resolution of these energy deposits is in principle described by Equation~\eqref{eq:ECAL_res}. However, when working on a small invariant mass interval around the $Z$~peak, it is possible to ignore the energy dependence of the resolution. Moreover, the left tail of the Gaussian resolution function is typically replaced with a more slowly decaying tail function in order to account for energy losses in the ECAL. It is therefore customary to model the smearing of the invariant mass by convolving the true intensity $f(m)$ with the so-called \emph{Crystal Ball \emph{(CB)} function} \citep{Oreglia1980, CMS2013ECAL}
\begin{equation}
 \mathrm{CB}(m|\Delta m, \sigma^2, \alpha, \gamma) = \begin{cases} C e^{-\frac{(m-\Delta m)^2}{2\sigma^2}}, & \frac{m-\Delta m}{\sigma} > -\alpha, \\
C \left( \frac{\gamma}{\alpha} \right)^\gamma e^{-\frac{\alpha^2}{2}} \left( \frac{\gamma}{\alpha} - \alpha - \frac{m - \Delta m}{\sigma} \right)^{-\gamma}, & \frac{m - \Delta m}{\sigma} \leq - \alpha, \end{cases}
\end{equation}
where $\sigma, \alpha, \gamma > 0$ and $C$ is a normalization constant chosen so that the function is a probability density. The Crystal Ball function is a Gaussian density with mean $\Delta m$ and variance $\sigma^2$ where the left tail is replaced with a power-law function. The parameter $\alpha$ controls the location of the transition from exponential decay into power-law decay and the parameter $\gamma$ controls the decay rate of the power-law tail.

The dataset we use is a digitized version of the lower left hand plot of Figure 11 in \citet{CMS2013ECAL}. These data correspond to an integrated luminosity\footnote{The number of particle reactions that took place in the accelerator is proportional to the integrated luminosity. As such, it is a measure of the amount of data produced by the accelerator. It is measured in the units of inverse femtobarns, $\mathrm{fb}^{-1}$.} of $4.98\ \mathrm{fb}^{-1}$ collected at the LHC in 2011 at the $7\ \mathrm{TeV}$ center-of-mass energy and include 67\:778 electron-positron events with the measured invariant mass between 65~GeV and 115~GeV. The data are discretized using a histogram with 100 bins of uniform width. The chosen electrons and positron have narrow particle showers in the central parts of the ECAL and as such correspond to ``high quality'' electron-positron pairs. For more details on the event selection, see \citet{CMS2013ECAL} and the references therein.

In order to estimate the parameters of the Crystal Ball function, we divided the dataset into two independent samples by drawing a binomial random variable independently for each bin with the number of trials equal to the observed bin contents. Consequently, the bins of the resulting two histograms are marginally mutually independent and Poisson distributed. Each observed event had a 70\:\% probability of belonging to the sample $\bm{y}$ used for unfolding and a 30\:\% probability of belonging to the sample used for CB parameter estimation.

The CB parameters $(\Delta m, \sigma^2, \alpha, \gamma)$ were estimated using maximum likelihood with the subsampled data on the full invariant mass range 65--115~GeV. The maximum likelihood estimates were
\begin{equation}
 (\Delta \hat{m}, \hat{\sigma}^2, \hat{\alpha}, \hat{\gamma}) = (0.58\ \mathrm{GeV}, (0.99\ \mathrm{GeV})^2, 1.81, 1.60)
\end{equation}
indicating that the measured invariant mass is on average 0.58 GeV too high and has an experimental resolution of approximately 1 GeV. As a cross-check of the fit, the estimated Crystal Ball function was used to smear the Breit--Wigner shape of the $Z$ boson invariant mass to obtain the corresponding expected smeared histogram, which was found to be in good agreement with the observations.

\subsection{Unfolding setup and results}

To carry out the empirical Bayes unfolding of the $Z$ boson invariant mass, we used the subsampled $n = 30$ bins on the interval $F = [82.5\ \mathrm{GeV},97.5\ \mathrm{GeV}]$. The resulting histogram $\bm{y}$ had a total of 42\:475 electron-positron events. To account for events that are smeared into the observed interval $F$ from the outside, we let the true space $E = [81.5\ \mathrm{GeV}, 98.5\ \mathrm{GeV}]$, that is, we extended it by approximately $1\hat{\sigma}$ on both sides with respect to $F$. The true space $E$ was discretized using order-4 B-splines with $L = 34$ uniformly placed interior knots resulting in $p = 38$ unknown spline coefficients. It was found out that such overparameterization with $p > n$ facilitated the mixing of the MCMC sampler. With these choices, the condition number of the smearing matrix was $\mathrm{cond}(\bm{K}) \approx 9.0 \cdot 10^3$. The boundary hyperparameters were set to $\gamma_\mathrm{L} = \gamma_\mathrm{R} = 70$.

The MCEM algorithm was initialized with $\delta^{(0)} = 1 \cdot 10^{-6}$ and was run for 20 iterations. During each MCEM iteration, the single-component Metropolis--Hastings algorithm was used to obtain 500 post-burn-in observations and the final point estimate $\bm{\hat{\beta}}$ was computed using a sample size of $5\:000$ observations. As above, the MCMC sampler was initialized with the non-negative least-squares fit to the smeared data. However, since $E \supsetneq F$, we extended $\bm{y}$ to match the size of $E$ by replicating the leftmost and the rightmost observations when computing the least squares fit. To form the bootstrap confidence intervals, $R = 200$ bootstrap replications were computed using resampling scheme 1. Running the MCEM iteration once to find the point estimate~$\bm{\hat{\beta}}$ took 5 minutes. With the bootstrap, the running time of the whole algorithm was 6 h 13 min with the bootstrap computations parallelized on the four cores.

The convergence of the MCEM algorithm was confirmed using a plot similar to Figure \ref{fig:gmm20000AlphaConv}. The algorithm converged in approximately 10 iterations to the hyperparameter estimate $\hat{\delta} = 7.4\cdot 10^{-8}$ with little Monte Carlo variation. During the MCEM iteration, the proposal acceptance rate remained at roughly $98\:\%$ and the average autocorrelation time increased from 3.0 to 8.6 indicating reasonable performance of the sampler throughout the whole iteration. As earlier, plots similar to Figure \ref{fig:gmm20000MCMC} were produced for each component of $\bm{\beta}$ for the final MCMC run in order to verify the appropriate convergence and mixing of the sampler.

\begin{figure}[!t]
\centering
%\vspace{6pc}
\includegraphics[trim = 0cm 0cm 0cm 0cm, clip=true, width=11.5cm]{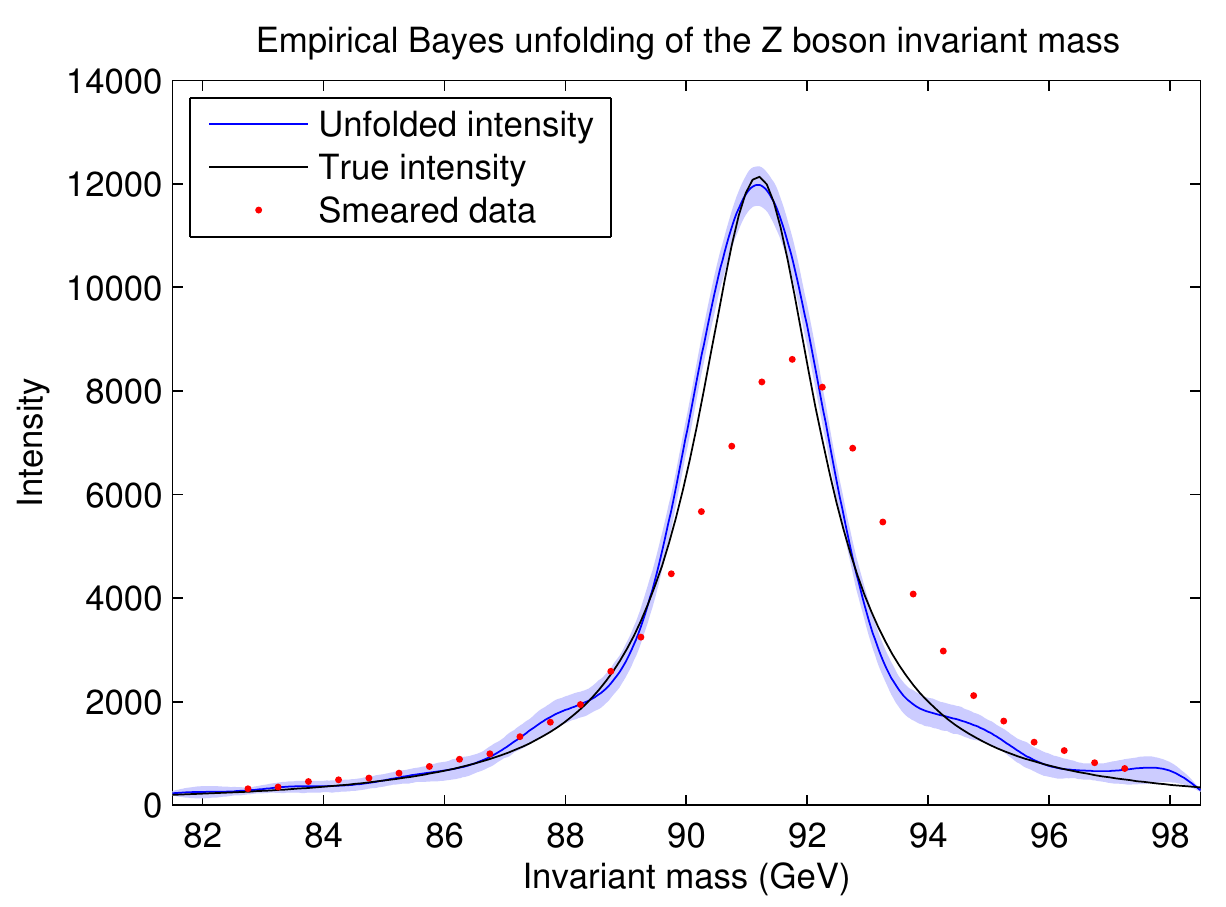}
\caption{Empirical Bayes unfolding of the $Z$ boson invariant mass spectrum. The unfolded intensity has been corrected for bias using bootstrap bias correction and the confidence band consists of 95\:\% pointwise basic bootstrap intervals. The red points show a histogram estimate of the smeared intensity.}
\label{fig:Zee}
\end{figure}

In Figure \ref{fig:Zee}, the bias-corrected unfolded intensity $\hat{f}_\mathrm{BC}$ of the $Z$ boson invariant mass, along with 95\:\% pointwise basic bootstrap intervals, is compared with the Breit--Wigner shape of the true mass peak. We observe that empirical Bayes unfolding captures reasonably well the overall shape of the Breit--Wigner distribution with few undesired artifacts. The figure also shows a histogram estimate of the smeared intensity given by the observed event counts $\bm{y}$ divided by the 0.5 GeV bin size. We see that the unfolding algorithm is able to correctly reconstruct the location and width of the $Z$ mass peak which are both distorted by the smearing in the ECAL. Moreover, thanks to the smoothness penalty and the Aristotelian boundary conditions, the intensity is estimated reasonably well in the 1 GeV regions in the tails of the intensity where no smeared observations were available.

\begin{figure}[ht]
\centering
%\vspace{6pc}
\subfigure{
\includegraphics[trim = 0cm 0cm 0cm 0cm, clip=true, width=6cm]{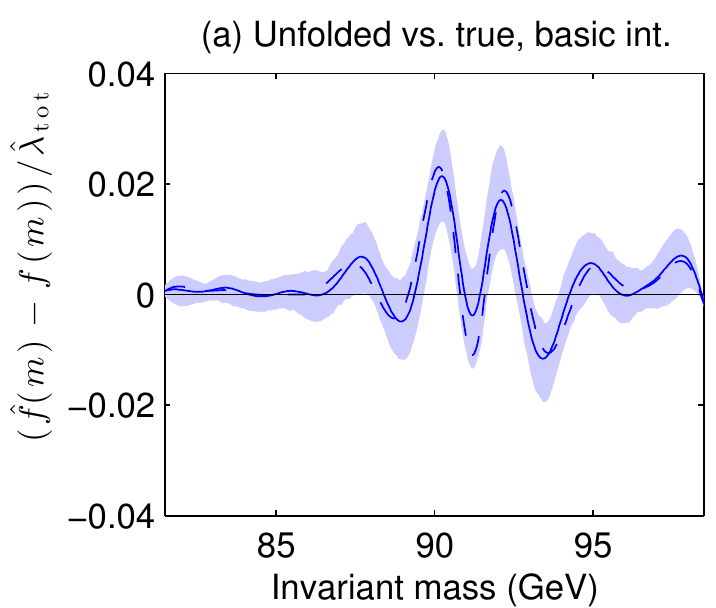}
\label{fig:ZeeDiffBasic}}
\subfigure{
\includegraphics[trim = 0cm 0cm 0cm 0cm, clip=true, width=6cm]{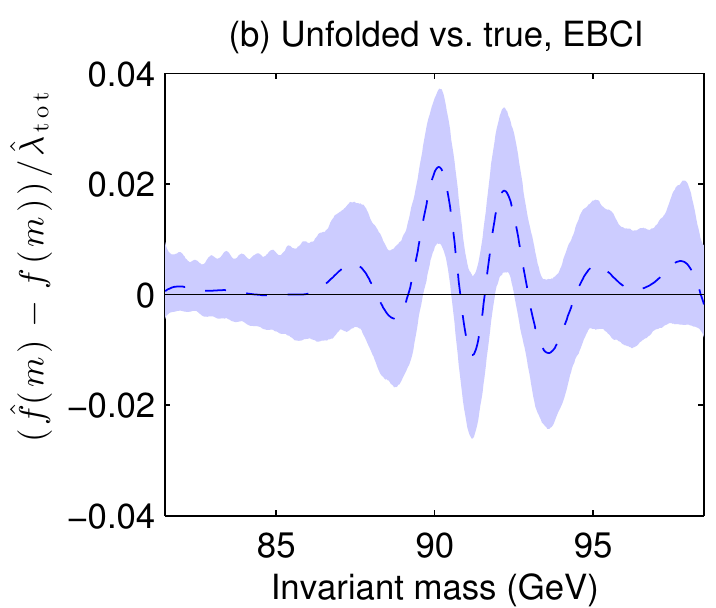}
\label{fig:ZeeDiffEBCI}}
\caption{Difference between the unfolded intensity $\hat{f}$ and the true intensity $f$ of the $Z$ boson invariant mass normalized for an estimate of the expected sample size $\hat{\lambda}_\mathrm{tot}$. Figure~(a) shows the 95\:\% pointwise basic bootstrap intervals, while Figure~(b) shows the corresponding na\"{i}ve empirical Bayes confidence intervals. Both figures include the original point estimate~$\hat{f}$ (dashed curve), and Figure (a) also shows the bias-corrected estimate~$\hat{f}_\mathrm{BC}$ (solid curve).}
\label{fig:ZeeDiff}
\end{figure}

However, in a closer examination, we observe that, starting from the top of the $Z$ mass peak, the unfolded intensity is first slightly too wide on both sides of peak and then slightly too narrow. This artifact is likely to be a residual bias which is not accounted for by the bootstrap bias correction. The end result of this effect is better seen in Figure \ref{fig:ZeeDiffBasic} which shows the unfolded intensity after subtraction of the true intensity $f$ and normalization for an estimate of the expected total number of events $\hat{\lambda}_\mathrm{tot} = \sum_{i=1}^n y_i$. The figure shows both the original point estimate $\hat{f}$ (dashed curve) and the bias-corrected estimate $\hat{f}_\mathrm{BC}$ (solid curve) along with the 95\:\% pointwise basic bootstrap intervals. Although it cannot be directly deduced from this figure, it seems likely that, because of the remaining bias, the confidence intervals do not attain their nominal 95\:\% frequentist coverage across the whole spectrum. See Section \ref{sec:discConc} for further discussion on this observation. Note also that the bias correction has improved the point estimate only at the top of the $Z$ boson mass peak but not at the sides of the peak.

To conclude this section, we show in Figure \ref{fig:ZeeDiffEBCI} the 95\:\% na\"{i}ve empirical Bayes confidence intervals for the $Z$ boson invariant mass. These intervals are again wider than the bootstrap intervals and hence seem to enjoy better coverage. Nevertheless, the interpretation of these intervals remains unclear.

\section{Concluding remarks} \label{sec:discConc}

We have studied a novel approach to solving the high energy physics unfolding problem involving empirical Bayes selection of the regularization strength and frequentist uncertainty quantification using the bootstrap. We have shown that empirical Bayes provides a principled way of choosing the hyperparameter $\delta$ with excellent practical performance in a wide variety of cases. As such, it provides an appealing alternative to classical methods for choosing the regularization strength, such as cross-validation or the Morozov discrepancy principle. Given the good performance of the approach, we anticipate empirical Bayes methods to also be valuable in solving other inverse problems beyond the unfolding problem.

It is nevertheless possible to find true intensities where empirical Bayes unfolding will not yield a good reconstruction. This happens when the smoothness penalty, i.e., penalizing for large values of $\|f''\|_2^2$, is not the appropriate way of regularizing the problem. For instance, if the true intensity $f$ contains sharp peaks or rapid oscillations, the solution would potentially be biased to an extent where the bootstrap bias correction would be unlikely to sufficiently alleviate the situation. Naturally, in such a case, a more suitable choice of the family of regularizing priors $\{p(\bm{\beta}|\delta)\}_{\delta > 0}$ should fix the situation. This highlights the fact that all the inferences considered here are contingent on the chosen family of priors and should always be interpreted with this in mind.

The other main component of our approach is frequentist uncertainty quantification of the solution using bootstrap resampling. We have shown that the bootstrap confidence intervals can serve as good estimates of the uncertainty of the solution, especially when there is little to moderate bias. However, with the $Z$ boson dataset studied in Section \ref{sec:Zboson}, it is likely that these intervals do not attain their nominal confidence level. There are several possible explanations for this. First, we did not take into account the uncertainty stemming from the estimation of the smearing matrix $\bm{K}$. Taking this uncertainty into account should widen the confidence intervals and hence improve coverage. The study of effective approaches to incorporating this uncertainty into the bootstrap procedure part of ongoing work. Second, the main problem in the unfolded $Z$ boson invariant mass shown in Figure~\ref{fig:Zee} is the presence of a bias in the form of small wiggles around the true intensity. The bootstrap is unable to probe this bias since the ill-posedness of~$\bm{K}$ ``smears away'' these oscillations when we compute the product $\bm{K}\bm{\hat{\beta}}$. In some sense, the bootstrap is blind to these artifacts and is hence unable to account for them either in the confidence intervals or the bias correction. To alleviate this problem, one could consider more elaborate bootstrap schemes. Perhaps one could, for example, sample the bootstrapped observations from the $\mathrm{Poisson}(\bm{K}'\bm{\hat{\beta}})$ distribution, where $\bm{K}'$ is a regularized version of $\bm{K}$.

Quite surprisingly, we found that in all our experiments the na\"{i}ve empirical Bayes confidence intervals were longer than the bootstrap intervals, even though the former do not take into account the uncertainty regarding the choice of the hyperparameter $\delta$. In practice, this is likely to mean that when there is little bias, the empirical Bayes intervals will overcover, but with larger bias, they might provide better coverage than the shorter bootstrap intervals. This means that the empirical Bayes intervals could also potentially serve as useful measures of uncertainty, especially since they are significantly cheaper to compute than the bootstrap intervals. Nevertheless, the indisputable advantage of the bootstrap intervals is that they enjoy a clear-cut frequentist interpretation, while the meaning of the empirical Bayes intervals is at best unclear. Interestingly, the recent theoretical results by \citet{Petrone2012} could potentially be used to prove the asymptotic coverage of the empirical Bayes intervals. Alternatively, one could perhaps use confidence distributions \citep{Xie2013} to form a link between the empirical Bayes posterior and frequentist coverage. Such results would significantly help to demystify the meaning of the empirical Bayes intervals, at least in the asymptotic sense.

\section*{Acknowledgements}

We wish to warmly thank Bob Cousins, Anthony Davison, Tommaso Dorigo, Louis Lyons and Mikko Voutilainen for insightful discussions and their encouragement in the course of this work.

\bibliographystyle{apalike}
\bibliography{references}

\end{document}